\documentclass[twocolumn]{svjour3}         

\usepackage{amsmath}
\usepackage{graphicx}
\usepackage{bm}
\usepackage{stmaryrd}

\usepackage{epstopdf}

\ifdefined\rd
\else
\newcommand{\rd}{\,\mathrm{d}}
\fi

\newcommand{\pfrac}[2]{\frac{\partial #1}{\partial #2}}

\newcommand{\V}[1]{\bm{#1}}

\ifdefined\sgn
\else
\newcommand{\sgn}[1]{\mbox{sgn$\left(\V{#1}\right)$}}
\fi

\newcommand{\wcal}{\mathcal{W}}
\newcommand{\ubar}{\bar{u}}

\newcommand{\ca}{\theta}

\newcommand{\ep}{\epsilon}

\usepackage{graphicx}
\usepackage{mathptmx}      
\usepackage{amsfonts, amssymb, amsbsy, amsmath}
\usepackage{subfig,pstricks,psfrag}
\usepackage{afterpage}
\usepackage[ruled,vlined]{algorithm2e}
\usepackage{bm}
\usepackage{color}
\smartqed  

\newcommand{\citep}[1]{{(\cite{#1})}}

\renewcommand{\vec}[1]{\mathbf{\bm{#1}}}
\renewcommand{\[}{\begin{equation}}
\renewcommand{\]}{\end{equation}}
\newcommand{\svec}[1]{\bm{#1}}

\graphicspath{{Images/},{Figures/},{FiguresMatlab/}}

\journalname{}
\begin{document}
\title{Closure Relations for Shallow Granular Flows from Particle Simulations}
\author{Thomas Weinhart$^{1,2}$ \and Anthony Thornton${}^{1,2,\dagger}$ \and Stefan Luding$^{1}$ \and Onno Bokhove$^{2}$}
\institute{
${}^{1}$ Dept. of Mechanical Engineering, Univ. of Twente, The Netherlands\\
${}^{2}$ Dept. of Applied Mathematics, Univ. of Twente, The Netherlands\\
${}^\dagger$ Numerical Analysis and Computational Mechanics group and MultiScale Mechanics group,
University of Twente, P.O. Box 217, 7500 AE  Enschede, The Netherlands,
Tel.: +31 53 489 3301, Fax: +31 53 489 4833,          
\email{a.r.thornton@utwente.nl}
}
\date{Received: date / Accepted: date}
\maketitle

%
%
%
\begin{abstract}
The Discrete Particle Method (DPM) is used to model granular flows down an inclined chute.
We observe three major regimes: static piles, steady uniform flows and accelerating flows.
For flows over a smooth base, other (quasi-steady) regimes are observed
where the flow is either highly energetic and strongly layered in depth for small inclinations,
or non-uniform and oscillating for larger inclinations.

For steady uniform flows, depth profiles of density, velocity and stress have been obtained using an improved coarse-graining method,
which allows accurate statistics even at the base of the flow.
A shallow-layer model for granular flows is completed with macro-scale closure relations
obtained from micro-scale DPM simulations of steady flows.
We thus obtain relations for the effective basal friction, shape factor, mean density,
and the normal stress anisotropy as functions of layer thickness, flow velocity and basal roughness.
For collisional flows, the functional dependencies are well determined and have been obtained.

\end{abstract}

\section{General introduction}

\subsection{Background}

Granular avalanche flows are common to natural environments and industry.
They occur across many orders of magnitude.
Examples range from rock slides, containing upwards of 1000$\,\mathrm{m}^3$ of material;
to the flow of sinter, pellets and coke into a blast furnace for iron-ore melting;
down to the flow of sand in an hour-glass.
The dynamics of these flows are influenced by many factors such as: polydisperity;
variations in density; non-uniform shape; complex basal topography; surface contact properties;
coexistence of static, steady and accelerating material; and, flow obstacles and constrictions.

Discrete Particle Methods (DPMs) are an extremely powerful way to investigate
the effects of these and other factors.
With the rapid recent improvement in computational power the full simulation of the
flow in a small hour glass of millions of particles is now feasible.
However, complete DPM simulations of large-scale geophysical mass flow will, probably, never be possible.

One primary goal of the present research is to simulate large scale and complex industrial flows using  granular shallow-layer equations. In this paper we will take the first step of using the DPM \cite{CundallStrack1979,SilbertErtasGrestHaleyLevinePlimpton2001,SilbertGrestPlimptonLandry2002,SilbertLandryGrest2003,Luding2008} to simulate small granular flows of mono-dispersed spherical particles in steady flow situations. We will use a refined and novel analysis to investigate three particular aspects of shallow chute flows:
\emph{i)} how to obtain meaningful macro-scale fields from the DPM simulation,
\emph{ii)} how to asses the flow dependence on the basal roughness, and
\emph{iii)} how to validate the assumptions made in depth-averaged theory.

The DPM simulations presented here will enable the construction of the mapping between the micro-scale
and macro-scale variables and functions, thus enabling construction of a closed set of continuum equations.
These mappings (closure relations) can then be used in continuum shal\-low-layer models and
compared with full DPM simulations (DPMs). For certain situations, precomputed closures should work;
but, in very complicated situations the pre-established relations may fail.
Heterogeneous, multi-scale modelling (HMM) is then an alternative
\cite{WeinanEngquistBjornLiRenVandern-Eijnden2007},
in which the local consititute relations are directly used in the continuum model.
In HMM, continuum and micro-scale models are dynamically coupled with a two-way communication
between the different models in selective regions in
both space and time, thus reducing computational expense and allowing simulation of complex granular flows.

%
\subsection{Shallow-layer models}

Shallow granular continuum models are often used to simulate geophysical mass flows,
including snow avalan\-ches \cite{CuiGrayJohannesson2007}, dense pyroclastic flows, debris flows \cite{DenlingerIverson2001}, block and ash flows \cite{DalbeyPatraPitmanBursikSheridan2008} and lahars \cite{WilliamsStintonSheridan2008}.
Such shallow-layer models involve approximations reducing the properties of a huge number of individual particles to a handful of averaged quantities.
Originally these models were derived from the general continuum incompressible mass and momentum equations,
using the long-wave approximation \cite{SavageHutter1989,HutterSiegelSavageNohguchi1993,GrayWielandHutter1999,WielandGrayHutter1999,GrayTaiNoelle2003,BokhoveThornton2010} for shallow variations in the flow height and basal topography.
Despite the massive reduction in degrees of freedom made,
shallow-layer models tend to be surprisingly accurate,
and are thus an effective tool in modelling geophysical flows.
Moreover, they are now used as a geological risk assessment and
hazard planning tool \cite{DalbeyPatraPitmanBursikSheridan2008}.
In addition to these geological applications, model applications involve
small-scale laboratory chute flows containing obstacles
\cite{GrayTaiNoelle2003}, wedges \cite{HakonardottirHogg2005,GrayCui2007} and
contractions \cite{VremanAlTaraziKuipersVanSintAnnalandBokhove2007},
showing good quantitative agreement between theory and experiment.

In fluid dynamics, the Navier-Stokes equations are established with full constitutive equations.
Nonetheless, the shallow-layer equations or Saint-Venant equations are often used 
in large scale situations where it is unpractical to solve the full Navier-Stokes equations.
Our present aim is to directly investigate the validity of the assumptions of granular shallow-layer
models first from discrete particle simulations, before obtaining fully 3D \lq kinetic theory\rq-style constitutive relations and
simplifying these via the depth-integration process.
A discussion of the full three-dimensional properties of our particle simulations will be undertaken later.
Here we restrict attention to the closures required for two-dimensional shallow granular flow equations. 

A key difference between shallow-layer fluid models and granular ones is the appearance of
a basal friction coefficient, $\mu$, being the ratio of the shear to normal traction at the base.
In early models, a dry Coulomb-like friction law was used \cite{SavageHutter1989}.
It implies $\mu$ to be constant, given by the tangent of the friction angle between the material
and the base, $\delta$, \emph{i.e.}, $\mu = \tan \delta$.
As a consequence constant uniform flow is only possible in such a model at that angle $\delta$,
independent of height.
There is a considerable amount of experimental evidence, {\em e.g.}, \cite{DaerrDouady1999,GDRMiDi2004},
that suggests that such a simple Coulomb law does not hold on rough beds or for moderate inclination angles. 
Furthermore, detailed experimental investigations using glass beads \cite{Pouliquen1999a} lead to an improved empirical `Pouliquen' friction law
characterised by two angles: the angle at which the material comes to rest, $\delta_1$,
where friction dominates over gravity and the angle, $\delta_2$, above which gravity dominates over friction and the material accelerates. Between these two angles steady flow is possible, and in the limit $\delta_1 \rightarrow \delta_2 = \delta$ the original Coulomb style model is recovered. 

Since its formulation a lot of work has been performed on extending and understanding this Pouliquen law.
The original law was obtained by retarding flowing material and measuring the angle at which the
material stopped as a function of height $h_{stop}\left( \theta \right)$, or equivalently,
by inverting this relation, $\theta_{stop}\left(h\right)$. For most materials, granular included,
a greater angle is required to initiate stationary than to retard flowing material.
Pouliquen and Forterre \cite{PouliquenForterre2002}, by measuring the angle required to start motion,
measured $\theta_{start}\left(h\right)$, \emph{i.e.}, the friction law for initially stationary material.
As expected $\theta_{start}$  was greater than $\theta_{stop}$ and this information was used to extend the
friction law to all values of the height and velocity within the steady regime.
Borzsonyi \& Ecke performed a series of additional experiments:
firstly, in \cite{BorzsonyiEcke2006} they looked at higher angles were the mean flow rates are close to the terminal velocity of a single particle, and found regions were the Pouliquen law is not valid and the Froude number becomes inversely proportional to the height, as opposed to the linear relationship observed for steady flow.
Borzsonyi and Ecke, and Pouliquen and Forterre \cite{BorzsonyiEcke2007,ForterrePouliquen2003} have all worked on extending the original law to be valid for more complicated non-spherical materials like sand and metallic materials.
Also, the effect of basal surface roughness has been systemically studied in \cite{GoujonThomasDalloz-Dubrujeaud2003} by varying the size of both the free flow and fixed basal particles.
For convenience, we define $\lambda$ to be the size ratio of the fixed and the free particles.
They observed a peak in roughness at a certain diameter ratio, $\lambda_c$,
which depends on the compactness of the basal layer.
Measured values of $\lambda_c$ in \cite{GoujonThomasDalloz-Dubrujeaud2003} ranged between $1$ and $3$ for a monolayer of fixed particles.
For fixed particles with smaller size and $\lambda<\lambda_c$, the range of angles where steady flow was observed decreased, and eventually the steady flow regime completely vanished, \emph{i.e.}, $\delta_2-\delta_1 \rightarrow 0$ as $\lambda \rightarrow 0$
(yielding Coulomb type behaviour).
For smaller flow particle diameters, {\em i.e.}, with $\lambda>\lambda_c$, there was also a reduction in friction, but weaker than in the small $\lambda$ case. For much larger $\lambda$, the friction saturated to a constant value, which they contributed to free particles that filled the holes in the basal surface and effectively created a stable basal surface of free particles. In a later publication \cite{GoujonDalloz-DubrujeaudThomas2007}, they extended this investigation to flows containing two particle sizes.

Louge and Keast \cite{LougeKeast2001} modified the kinetic theory presented in \cite{Jenkins1993} by modelling enduring contacts
via a frictional rate-independent stress component in order to obtain steady flow on flat frictional inclines. This work was later extended to bumpy inclines \cite{Louge2003}. Jenkins \cite{Jenkins2006} took a different approach and theoretically formulated a phe\-no\-me\-no\-lo\-gi\-cal modification of granular kinetic theory to account for enduring particle contacts. His idea is that enduring contacts between grains, forced by the shearing, reduce the collision rate of dissipation. Therefore a modification to the dissipation is introduced, which does not affect the stress.
It leads to a law very similar to the one experimentally obtained by Pouliquen.
He extended the theory in \cite{Jenkins2007} to very dissipative frictional particles, with a coefficient of restitution less than $0.7$.
Later, a detailed comparison with new experiments was performed, showing agreement with flows of low inclination \cite{JenkinsBerzi2010}. 

\newcommand{\gapprox}{\hbox{\lower .8ex\hbox{$\,\buildrel > \over\sim\,$}}}

Silbert \emph{et al.} \cite{SilbertErtasGrestHaleyLevinePlimpton2001} used DPMs to simulated chute flow of cohesionless particles.  They found that a steady-state flow regime exists over a wide range of inclination angles, heights and interaction parameters, in confirmation of the experiments of Pouliquen \cite{Pouliquen1999a}.  They found for steady-state flows that the volume fraction is constant throughout the flow, in agreement with the assumptions of shallow-layer theory \cite{SavageHutter1989}.  They also observed that the shear stress is proportional to the square of the shear and the flow velocity scales with the height to the power $3/2$.
This result coincides with Bagnold's analysis of dilute binary collisions flows \cite{Bagnold1954}.
They also observed small systematic deviations from isotropic stress, which shows a deviation from fluid-like behaviour.  However, normal stresses do not approach a Coulomb-yield criterion structure at the angle of repose except near the surface, hinting that the failure of flow starts near the surface.  
In \cite{SilbertGrestPlimptonLandry2002}, they looked at the effect of different basal types and found that for an ordered chute base the steady state regime splits into three distinct flow regimes: at smaller angles, the flowing system self-organizes into a state of low-dissipation flow consisting of in-plane ordering in the bulk; at higher angles, a high-dissipation regime similar to that for a rough base but with considerable slip at the bottom is observed; and, between these two sub-regions they observe a transitional flow regime characterized by large oscillations in the bulk averaged kinetic energy due to the spontaneous ordering and disordering of the system as a function of time.
Finally, \cite{SilbertLandryGrest2003} investigated the initiation and cessation of granular chute flow more carefully and
computed both $\theta_{stop}$ and $\theta_{start}$. For inclinations $\theta\gg\theta_{stop}$ they observed a Bagnold rheology, for $\theta\gapprox\theta_{stop}$ a linear profile, and for $\theta\approx\theta_{stop}$ avalanching behaviour.

\subsection{Overview of this study}

Our present research is novel on the following three counts:

Firstly, we compute more meaningful macro-scale fields from the DPM simulations than before
by carefully choosing the coarse graining function.
In order to homogenize the DPM data, the micro-scale fields need to be coarse-grained to obtain macroscopic fields.
Coarse-grained micro-scale fields of density,
momentum and stress were derived directly from the mass and momentum balance
equations by Goldhirsch \cite{Goldhirsch2010}. 
The quality of the statistics involved depends on the coarse graining width $w$,
which defines the amount of spatial smoothing. 
For small coarse-graining width $w$, micro-scale variations remain visible,
while for large $w$ these smooth out in the macro-scale gradients.
Since one of the objectives is to obtain the value of $\mu$ at the base,
we use a novel adaptation of Goldhirsch' statistics \cite{Goldhirsch2010} near boundaries.
This is a non-trivial issue neglected in the literature; often continuum fields
are simply not computed within a distance $w$ of the boundary.
Our new approach \cite{WeinhartThorntonBokhoveLuding2011} is consistent with the continuum equations everywhere,
enabling the construction of continuum fields within one course-graining width of the boundary. 

Secondly, we follow the approach of \cite{GoujonThomasDalloz-Dubrujeaud2003}
and vary the basal particle diameter to achieve different basal conditions.
For particles with smaller basal than flowing diameter, $\lambda <1$, the flow becomes more energetic and
oscillatory behaviour is observed. This phenomena has previously been reported in \cite{SilbertGrestPlimptonLandry2002}, but was achieved by changing the basal particles to a more regular, grid-like configuration.
By investigating flow over fixed particles of different size than the free ones, we are able to quantify the roughness
and numerically investigate the transition from rough to smooth surfaces.
For smoother surfaces, we show that the parameter space can be split into to two types of steady flow,
and we obtain a general friction law. 

Finally, we test the assumptions made in depth-ave\-rag\-ed theory and determine the required closure laws.
For shallow granular flows, the flow can be described by depth-ave\-rag\-ed mass and momentum-balance equations \cite{GrayTaiNoelle2003}.
Solving the depth-averaged equations requires a constitutive relation for the basal friction,
a way to account for mean density variations, the shape of the velocity profile and the pressure anisotropy.
We extract such data from DPMs obtained for steady uniform flows,
and establish a novel, extended set of closure equations.
Also, the depth-averaged equations are obtained under the assumptions that
a) the density is constant in space and time and does not vary through the flow;
b) the ratio between mean squared velocity and the squared mean velocity is known;
c) the downward normal stress is lithostatic, \emph{i.e.},
balances the gravitational forces acting on the flow;
and, d) the ratio between the normal stresses is known.
Gray {\it et al.} \cite{GrayTaiNoelle2003} assumed the latter ratio to be one.
The depth profiles of these quantities are discussed by Silbert {\it et al.} in \cite{SilbertErtasGrestHaleyLevinePlimpton2001,SilbertGrestPlimptonLandry2002,SilbertLandryGrest2003} for steady flow.
We extend these measurements to validate our DPM simulations, using our superior statistical procedure.
Hence, we establish the range in which the shallow-layer model is valid,
and the closure relations required for shallow-layer continuum simulations,
for a much wider range of flow regimes than had been considered before.

\subsection{Outline}

We introduce the force model used in the DPM in Sect.~\ref{sec:contactlaw}, and the statistical method
used to obtain macroscopic density, velocity and stress profiles in Sect.~\ref{sec:statistics}.
In Sect.~\ref{sec:math}, we discuss the continuum shallow-layer equations for modelling granular flow
including some macro-scale closures.
The set up of the simulations is discussed in Sect.~\ref{sec:simulation},
and the steady-state regime is mapped for flows over a rough basal surface in Sect.~\ref{sec:rough}.
Depth profiles of the flow are introduced in Sect.~\ref{sec:w},
which are then used to characterize the steady flow over smoother surfaces in Sect.~\ref{sec:smooth}.
Finally, the closure relations for the shallow-layer model are established in Sect.~\ref{sec:closure},
before we conclude in Sect.~\ref{sec:conclusion}.

\section{Contact law description}\label{sec:contactlaw}

%
A Discrete Particle Method (DPM) is used to perform the simulation of a collection of identical granular particles.
Boundaries are created by special fixed particles, which generally will have different properties than the flow particles.
Particles interact by the standard spring-dash\-pot interaction model \cite{CundallStrack1979,WaltonBraun1986,Luding2008},
in which it is assumed that particles are spherical and soft, and that pairs have a single contact point.

Each particle $i$ has a diameter $d_i$, density $\rho_i$, position $\vec{r}_i$, velocity $\vec{v}_i$
and angular velocity $\svec{\omega}_i$.
For pairs of two particles $\{i,j\}$, we define the relative distance vector $\vec{r}_{ij}=\vec{r}_i-\vec{r}_j$,
their separation $r_{ij} = |\vec{r}_{ij}|$, the unit normal
$\hat{\vec{n}}_{ij} = \vec{r}_{ij}/r_{ij}$ and the relative velocity $\vec{v}_{ij}=\vec{v}_i-\vec{v}_j$.
Two particles are in contact if their overlap,
\begin{equation}
\delta_{ij}^n=\max(0,(d_i+d_j)/2-r_{ij}), \label{eq:deltan}
\end{equation}
is positive. A single contact point $\vec{c}$ at the centre of the overlap is assumed,
which is a valid assumption as long as the overlap is small.
For our simulations the overlap between two particles is always below $1\%$
of the particle radius; the simple contact model we employ thus captures the key process
as there are no multiple contact points.



The force acting on particle $i$ is a combination of the pairwise interaction of two particles. 
The force $\vec{f}_{ij}$ represents the force on particle $i$ from the interaction with
particle $j$ and can be decomposed into a normal and a tangential component,
\begin{equation}
\vec{f}_{ij} = \vec{f}_{ij}^{\,n} + \vec{f}_{ij}^{\,t}.
\end{equation}

We assume that the particles are viscoelastic; therefore, they experience elastic as well as
dissipative forces in both normal and tangential directions.
The normal force is modelled as a spring-dashpot with a linear elastic and a linear dissipative contribution,
\begin{equation} \vec{f}_{ij}^{\,n}  = k^n \delta_{ij}^{\,n} \hat{\vec{n}}_{ij} -\gamma^{\,n} \vec{v}_{ij}^n,
\end{equation}
with spring constant $k^n$, damping coefficient $\gamma^n$ and the normal relative velocity component,
\begin{equation} 
\vec{v}_{ij}^n = (\vec{v}_{ij} \cdot \hat{\vec{n}}_{ij}) \hat{\vec{n}}_{ij}.
\end{equation} 
For a central collision, no tangential forces are present, and the collision time $t_c$ between two particles can be calculated as
\begin{equation}\label{eq:tc}
t_c=\pi/\sqrt{\frac{k^n}{m_{ij}}-\left(\frac{\gamma^{\,n}}{2m_{ij}}\right)^2},
\end{equation} 
with the reduced mass $m_{ij}=m_i m_j/(m_i+m_j)$.
The normal restitution coefficient $r_c$ (ratio of relative normal speed after and before collision) is calculated as
\begin{equation}
r_c = \exp(-t_c \gamma^{\,n}/(2m_{ij})).
\end{equation}

We also assume a linear elastic and a linear dissipative force in the tangential direction,
\begin{equation} 
\vec{f}_{ij}^{\,t} = -k^{\,t} \pmb{\delta}_{ij}^t -\gamma^{\,t} \vec{v}_{ij}^t,\end{equation}
with spring constant $k^{\,t}$, damping coefficient $\gamma^{\,t}$,
elastic tangential displacement $\pmb{\delta}_{ij}^t$ (which is explained later), and the relative tangential velocity at the contact point,
\begin{equation} \label{eq:vt}
\vec{v}_{ij}^t = \vec{v}_{ij} - \vec{v}_{ij}^n - \svec{\omega}_{i}\times\vec{b}_{ij}+\svec{\omega}_{j}\times\vec{b}_{ji}, 
\end{equation} 
with $\vec{b}_{ij}=-\bigl((d_{i}-\delta_{ij}^n)/2\bigr)\hat{\vec{n}}_{ij}$
the branch vector from point $i$ to the contact point;
for equal size particles $\vec{b}_{ij}=-\vec{r}_{ij}/2$.



The elastic tangential force is modelled and derives from the roughness of the particle surface.
Near the contact point, small bumps on a real particle would stick to each other, due to the normal force pressing them together,
and elongate in the tangential direction resulting in an elastic force proportional to the elastic tangential displacement.
It is defined to be zero at the initial time of contact, and its rate of change is given by
\begin{equation}\label{eq:deltat}
\frac{d \pmb{\delta}_{ij}^t}{dt} =
\vec{v}_{ij}^t-\frac{(\pmb{\delta}_{ij}^t\cdot \vec{v}_{ij})\hat{\vec{n}}_{ij}}{r_{ij}},
\end{equation}
where the first term is the relative tangential velocity at the contact point,
and the second term ensures that $\pmb{\delta}_{ij}^t$ remains normal to $\hat{\vec{n}}_{ij}$.
The second term is always orthogonal to the spring direction and, hence, does not affect the rate of
change of the spring length: it simply rotates it, thus keeping it tangential.

When the tangential to normal force ratio becomes larger than the particle
contact friction coefficient, $\mu_c$, for a real particle
the bumps would slip against each other and their elongation is shortened until the
bumps can stick to each other again.
This is modelled by a static yield criterion, truncating the magnitude of $\pmb{\delta}_{ij}^t$ as necessary
to satisfy $|\vec{f}_{ij}^{\,t}| \le \mu_c |\vec{f}_{ij}^{,n}|$.
Thus, the contact surfaces are treated as stuck while $|\vec{f}_{ij}^{\,t}|<\mu_c |\vec{f}_{ij}^{\,n}|$ and
as slipping otherwise, when the yield criterion is satisfied\footnote{{\em Meant for review stage only.}
It should be noted that in the absence of dissipative forces and slipping,
the system can be described as an Hamiltonian system: see Appendix \ref{sec:hamiltonian}.
Appendix \ref{sec:tangentialspring} contains details on the tangential displacement.
A pseudocode of the tangential force calculation is provided in  Appendix \ref{sec:ft}.}.

The total force on particle $i$ is a combination of contact forces $\vec{f}_{ij}$ with other particles and external
forces such as gravity $\vec{g}$. The resulting force $\vec{f}_i$ and torque $\vec{q}_i$ acting on particle $i$ are
\begin{equation} 
\vec{f}_i = \vec{g} + \sum_{j=1,j\not=i}^N \vec{f}_{ij},
\quad \mbox{and}\quad
\vec{q}_i = \sum_{j=1,j\not=i}^N \vec{b}_{ij} \times \vec{f}_{ij}.
\end{equation}
Finally, using these expressions we arrive at Newton's equations of motion for the translational and rotational degrees of freedom,
\begin{equation}
m_i \frac{{\rm d}^2 \vec{r}_i}{{\rm d} t^2}  = \vec{f}_i,
\quad \mbox{and}\quad
I_i \frac{{\rm d}}{{\rm d} t} \svec{\omega}_i = \vec{q}_i,
\label{eq:newton}
\end{equation}
with $m_i$ the mass and $I_i$ the inertia of particle $i$. 
We integrate \eqref{eq:newton} forward using Velocity-Verlet \cite{AllenTildesley1993},
formally second order in time,
with an adequate time step of $\Delta t=t_c/50$. The collision time $t_c$ is given by \eqref{eq:tc},
while \eqref{eq:deltat} is integrated using first-order forward Euler.

Hereafter, we distinguish between identical free flowing and identical fixed basal particles. Base particles are modelled as having an infinite mass and are unaffected by body forces: they do not move.
This leaves two distinct types of collision: flow-flow, and flow-base. Model parameters for each of these collision types are changed independently.

\section{Statistics}\label{sec:statistics}
\subsection{Coarse-graining}\label{s:sd:1}

The main aims of this paper are to use discrete particle simulations to both confirm 
the assumptions of and provide closure rules required for the depth-averaged shallow-water equations.
Hence, continuum fields have to be extracted from the discrete particle data.
There are many papers in the literature on how to go from the discrete to the continuum,
binning micro-scale fields into small volumes
\cite{IrvingKirkwood1950,SchofieldHenderson1982,Luding2004,Luding2009,LudingLatzelVolkDiebelsHerrmann2001},
averaging along planes \cite{ToddEvansDaivis1995},
or coarse graining spatially and temporally \cite{Babic1997,ShenAtluri2004,Goldhirsch2010}.
Here, this will be achieved using a coarse-graining approach first described in \cite{Babic1997}
based on a later derivation of Goldhirsch \cite{Goldhirsch2010},
extended further by us \cite{WeinhartThorntonBokhoveLuding2011} to account for boundary forces
due to the presence of the base. 

The method has several advantages over other methods because:
(i) the fields produced automatically satisfy the equations of continuum mechanics, even near the flow base;
(ii) it is neither assumed that the particles are rigid nor spherical; and,
(iii) the results are even valid for single particles as no averaging over groups of particles is required.
The only assumptions are that each particle pair has a single point of contact (\emph{i.e.},
the particle shapes are convex), the contact area can be replaced by a contact point (\emph{i.e.},
the particles are not too soft), and that collisions are not instantaneous. 

\subsection{Mass and momentum balance}\label{s:sd:2}
\subsubsection{Notation and basic ideas}

Vectorial and tensorial components are denoted by Greek letters in order to distinguish them from the Latin particle indices $i,j$.
Bold vector notation will be used when convenient.

Assume a system given by $N_p$ flowing particles and $N_b$ fixed particles.
Since we are interested in the flow, we will calculate macroscopic fields pertaining to the flowing particles only.
From statistical mechanics, the microscopic mass density of the flow,
$\rho^{\mbox{mic}}$, at a point $\V{r}$ at time $t$ is defined by 
\begin{equation}\label{eq:sd:1}
\rho^{\mbox{mic}} (\V{r},t) = \sum_{i=1}^{N_p} m_i \delta \left(\V{r} -\V{r}_i(t) \right),
\end{equation}
where $\delta(\V{r})$ is the Dirac delta function and $m_i$ is the mass of particle $i$.
The following definition of the macroscopic density of the flow is used
\begin{equation}\label{eq:sd:2}
\rho (\V{r},t) = \sum_{i=1}^{N_p} m_i \wcal \left(\V{r} -\V{r}_i(t)\right),
\end{equation}
thus replacing the Dirac delta function in \eqref{eq:sd:1} by an integrable
\lq coarse-graining\rq\ function $\wcal$ whose integral over space is unity.
We will take the coarse-graining function to be a Gaussian
\begin{equation}\label{eq:sd:3}
\wcal\left(\V{r} -\V{r}_i(t)\right) = \frac{1}{(\sqrt{2\pi} w)^3} 
\exp\left(-\frac{|\V{r}-\V{r}_i(t)|^2}{2 w^2}\right)
\end{equation}
with width or variance $w$.
Other choices of the coarse-graining function are possible, but the Gaussian has the advantage that it
produces smooth fields and the required integrals can be analyzed exactly. 
According to \cite{Goldhirsch2010}, the answer depends only
weakly on the choice of function, and the width $w$ is the key parameter.

It is clear that as $w \rightarrow 0$ the macroscopic density defined in \eqref{eq:sd:3} reduces to the one in \eqref{eq:sd:2}.
The coarse-graining function can also be seen as a convolution integral between the micro and macro definitions, 
\begin{equation}\label{eq:sd:4}
\rho (\V{r},t) = \int \wcal(\V{r}-\V{r'}) \rho^{\mbox{mic}} (\V{r}',t) \rd \V{r}'.
\end{equation}

\subsubsection{Mass balance}

Next we will consider how to obtain the other fields of interest:
the momentum vector field and the stress tensor. As stated in Sect.~\ref{s:sd:1} the macroscopic variables
will be defined in a way compatible with the continuum conservation laws. 

The coarse grained momentum density $\V{p}(\V{r},t)$ is defined by
\begin{equation}\label{eq:sd:5}
p_{\alpha}(\V{r},t) = \sum_{i=1}^{N_p} m_i v_{i \alpha} \wcal (\V{r} - \V{r}_i),
\end{equation}
where the ${v}_{i \alpha}$'s are the velocity components of particle $i$.
The macroscopic velocity field $\V{V}(\V{r},t)$ is then defined as the ratio of momentum and density fields, 
\begin{equation}\label{eq:sd:6}
V_{\alpha}(\V{r},t) = p_{\alpha}(\V{r},t)/\rho(\V{r},t).
\end{equation}
It is straightforward to confirm that equations \eqref{eq:sd:2} and \eqref{eq:sd:5} lead to the continuity equation
\begin{equation}\label{eq:mass}
\pfrac{\rho}{t} + \pfrac{p_{\alpha}}{r_{\alpha}} = 0,
\end{equation}
with the Einstein summation convention for Greek letters.

\subsubsection{Momentum balance}

Finally, we will consider the momentum conservation equation with the aim of establishing
the macroscopic stress field. In general, the desired momentum balance equations are written as,
\begin{equation}\label{eq:momentum}
\pfrac{p_{\alpha}}{t} = - \pfrac{}{r_{\beta}}\left[\rho V_{\alpha} V_{\beta}\right] + \pfrac{\sigma_{\alpha \beta}}{r_{\beta}} + \rho g_\alpha,
\end{equation}
where $\sigma_{\alpha\beta}$ is the stress tensor,
and $g_{\alpha}$ a component of the acceleration vector of gravity.

Expressions \eqref{eq:sd:5} and \eqref{eq:sd:6} for the momentum $\V{p}$ and the velocity $\V{V}$ have already
been defined. The next step is to compute their temporal and spatial derivatives, respectively,
and reach closure. Taking the time derivative of \eqref{eq:sd:5} gives
\begin{eqnarray}\label{eq:sd:9}
\pfrac{p_{\alpha}}{t} &=& \pfrac{}{t} \sum_{i=1}^{N_p} m_i v_{i \alpha}(t) \wcal(\V{r} -\V{r}_i(t)) \nonumber\\
&=& \sum_{i=1}^{N_p} m_i \dot{v}_{i \alpha} \wcal(\V{r} -\V{r}_i)  + \sum_{i=1}^{N_p} m_i v_{i \alpha} \pfrac{}{t}\wcal(\V{r} -\V{r}_i).
\end{eqnarray}
Using \eqref{eq:newton}, the first term in \eqref{eq:sd:9} can be expressed as
\begin{equation}\label{eq:sd:10}
A_{\alpha} \equiv \sum_{i=1}^{N_p} m_i \dot{v}_{i \alpha} \wcal(\V{r} -\V{r}_i) = \sum_{i=1}^{N_p} f_{i \alpha} \wcal(\V{r} -\V{r}_i). 
\end{equation}
In the simulations presented later the force on each particle contains three contributions:
particle-particle interactions, particle-base interactions, and the gravitational body force. 
Hence,
\begin{equation}\label{eq:sd:11}
f_{i \alpha} = \sum_{j=1, j \ne i}^{N_p} f_{ij \alpha} + \sum_{k=1}^{N_b} f_{ik \alpha}^\text{b} + m_i g_{\alpha},
\end{equation}
where $f_{ij}$ is the interaction force between particle $i$ and $j$,
and $f_{ik}^\text{b}$ the interaction between particle $i$ and base particle $k$, 
or base wall if the base is flat.
Therefore, we rework \eqref{eq:sd:10} as
\begin{equation}\label{eq:sd:12}
A_{\alpha} = \sum_{i=1}^{N_p} \sum_{j=1, j \ne i}^{N_p} f_{ij \alpha} \wcal_i + \sum_{i=1}^{N_p} \sum_{k=1}^{N_b} f_{ik \alpha}^\text{b}\wcal_i + \sum_{i=1} m_i \wcal_i g_{\alpha},
\end{equation}
where $\wcal_i = \wcal(\V{r} - \V{r}_i)$.
The last term in \eqref{eq:sd:12} can be simplified to $\rho g_{\alpha}$ by using \eqref{eq:sd:2}. From Newton's third law, the contact forces are equal and opposite, such that $f_{ij}=-f_{ji}$.Hence,
\begin{equation}\label{eq:sd:13}
\sum_{i=1}^{N_p} \sum_{j=1, j \ne i}^{N_p} f_{ij \alpha} \wcal_i = \sum_{i=1}^{N_p} \sum_{j=1, i \ne j}^{N_p} f_{ji \alpha} \wcal_j = - \sum_{i=1}^{N_p} \sum_{j=1, i \ne j}^{N_p} f_{ij \alpha} \wcal_j.
\end{equation}
where in the first step we interchanged the dummy summation indices.
It follows from \eqref{eq:sd:13} that \eqref{eq:sd:12} can be written as 
\begin{eqnarray}\label{eq:sd:14}
A_{\alpha} &=& \frac{1}{2}\sum_{i=1}^{N_p} \sum_{j=1, j \ne i}^{N_p} f_{ij \alpha} \left(\wcal_i - \wcal_j\right) + \sum_{i=1}^{N_p} \sum_{k=1}^{N_b} f_{ik \alpha}^\text{b}\wcal_i + \rho g_{\alpha} \nonumber\\ 
&=& \sum_{i=1}^{N_p} \sum_{j=i+1}^{N_p} f_{ij \alpha} \left(\wcal_i - \wcal_j\right) + \sum_{i=1}^{N_p} \sum_{k=1}^{N_b} f_{ik \alpha}^\text{b}\wcal_i + \rho g_{\alpha}.
\end{eqnarray}

Next, we will write $A_{\alpha}$ as the divergence of a tensor in order to obtain a formula for the stress tensor.
The following identity holds for any smooth function $\wcal$
\begin{eqnarray}\label{eq:sd:15}
\wcal_j - \wcal_i &=& \int_0^1 \pfrac{}{s} \wcal (\V{r} - \V{r}_i + s \V{r}_{ij}) \rd s \nonumber\\ 
&=&r_{ij \beta} \pfrac{}{r_{\beta}} \int_0^1  \wcal (\V{r} - \V{r}_i + s \V{r}_{ij}) \rd s,
\end{eqnarray}
where $\V{r}_{ij}=\V{r}_i-\V{r}_j$; we used the chain rule and differentiation to the full argument of
$\wcal(\cdot)$ per component.

The next step extends Goldhirsch' analysis near a boundary.
To obtain a similar expression for the interaction with base particles, we write
\begin{eqnarray}\label{eq:sd:15b}
-\wcal_i &=& \int_0^\infty \pfrac{}{s} \wcal (\V{r} - \V{r}_i + s \V{r}_{ik}) \rd s \nonumber\\ 
&=&r_{ik \beta} \pfrac{}{r_{\beta}} \int_0^\infty  \wcal (\V{r} - \V{r}_i + s \V{r}_{ik}) \rd s,
\end{eqnarray}
which holds because $\wcal_i$ decays towards infinity.
Substituting identities \eqref{eq:sd:15}, \eqref{eq:sd:15b} and \eqref{eq:sd:2} into \eqref{eq:sd:14} leads to
\begin{multline}\label{eq:sd:16}
A_{\alpha} = 
-\pfrac{}{r_{\beta}}  \sum_{i=1}^{N_p} \sum_{j=i+1}^{N_p} f_{ij \alpha} r_{ij \beta} \int_0^1  \wcal(\V{r} - \V{r}_i + s \V{r}_{ij} ) \rd s \\
- \pfrac{}{r_{\beta}} \sum_{i=1}^{N_p} \sum_{k=1}^{N_b} f_{ik \alpha}^\text{b} r_{ik \beta} \int_0^\infty  \wcal(\V{r} - \V{r}_i + s \V{r}_{ik} )\rd s + \rho g_{\alpha}.
\end{multline}
From \cite{Goldhirsch2010}, it follows that the second term in \eqref{eq:sd:9} can be expressed as follows
\begin{equation}\label{eq:sd:16b}
\sum_i m_i v_{i \alpha} \pfrac{}{t}\wcal(\V{r} -\V{r}_i) = 
-\pfrac{}{r_{\beta}} \left[ \rho V_{\alpha} V_{\beta} + \sum_i^{N_p} m_i v_{i \alpha}' v'_{i \beta} \wcal_i \right],
\end{equation}
where $\V{v}'_i$ is the fluctuating velocity of particle $i$, with components given by
\begin{equation}
v_{i \alpha}' (t,\V{r}) =v_{i \alpha} (t) - V_{\alpha} (\V{r},t).
\end{equation}
Substituting \eqref{eq:sd:16} and \eqref{eq:sd:16b} into momentum balance \eqref{eq:momentum} yields
\begin{multline}
\pfrac{\sigma_{\alpha \beta}}{r_{\beta}} = \pfrac{}{r_{\beta}} \left[-\sum_{i=1}^{N_p} \sum_{j=i+1}^{N_p} f_{ij \alpha} r_{ij \beta} \int_0^1  \wcal(\V{r} - \V{r}_i + s \V{r}_{ij} ) \rd s  \right.\\ 
- \sum_{i=1}^{N_p} \sum_{k=1}^{N_b} f_{ik \alpha}^\text{b} r_{ik \beta} \int_0^\infty  \wcal(\V{r} - \V{r}_i + s \V{r}_{ij} ) \rd s
\left.-\sum_i^{N_p} m_i v_{i \alpha}' v'_{i \beta} \wcal_i \right].
\end{multline}
Therefore the stress is given by
\begin{multline} \label{eq:sd:18}
\sigma_{\alpha \beta} = -\sum_{i=1}^{N_p} \sum_{j=i+1}^{N_p} f_{ij \alpha} r_{ij \beta} \int_0^1  \wcal(\V{r} - \V{r}_i + s \V{r}_{ij} ) \rd s\\ 
- \sum_{i=1}^{N_p} \sum_{k=1}^{N_b} f_{ik \alpha}^\text{b} r_{ik \beta} \int_0^\infty  \wcal(\V{r} - \V{r}_i + s \V{r}_{ik} ) \rd s
-\sum_i^{N_p} m_i v_{i \alpha}' v'_{i \beta} \wcal_i.
\end{multline}
The kinetic component of the stress tensor as well as the contact stress coming from normal forces are symmetric;
only the contact stress from tangential forces can contribute to the antisymmetric part of the stress tensor.
In our simulations the tangential forces contribute less than $5\%$ to the total stress in the system,
such that the stress is almost symmetric.

\begin{figure}[tbp]
%
\begin{psfrags}%
\psfragscanon%
%
\psfrag{s01}[t][t]{\fontsize{10}{15}\fontseries{m}\mathversion{normal}\fontshape{n}\selectfont \color[rgb]{0,0,0}\setlength{\tabcolsep}{0pt}\begin{tabular}{c}$x$\end{tabular}}%
\psfrag{s05}[l][l]{\fontsize{10}{15}\fontseries{m}\mathversion{normal}\fontshape{n}\selectfont \color[rgb]{0,0,0}$\partial_x \sigma$}%
\psfrag{s06}[l][l]{\fontsize{10}{15}\fontseries{m}\mathversion{normal}\fontshape{n}\selectfont \color[rgb]{0,0,0}$\rho$}%
\psfrag{s07}[l][l]{\fontsize{10}{15}\fontseries{m}\mathversion{normal}\fontshape{n}\selectfont \color[rgb]{0,0,0}$\sigma$}%
\psfrag{s08}[l][l]{\fontsize{10}{15}\fontseries{m}\mathversion{normal}\fontshape{n}\selectfont \color[rgb]{0,0,0}$\partial_x \sigma$}%
\psfrag{s10}[][]{\fontsize{10}{15}\fontseries{m}\mathversion{normal}\fontshape{n}\selectfont \color[rgb]{0,0,0}\setlength{\tabcolsep}{0pt}\begin{tabular}{c} \end{tabular}}%
\psfrag{s11}[][]{\fontsize{10}{15}\fontseries{m}\mathversion{normal}\fontshape{n}\selectfont \color[rgb]{0,0,0}\setlength{\tabcolsep}{0pt}\begin{tabular}{c} \end{tabular}}%
%
\fontsize{10}{15}\fontseries{m}\mathversion{normal}%
\fontshape{n}\selectfont%
%
\psfrag{x01}[t][t]{-2}%
\psfrag{x02}[t][t]{-1}%
\psfrag{x03}[t][t]{0}%
\psfrag{x04}[t][t]{1}%
\psfrag{x05}[t][t]{2}%
%
\psfrag{v01}[r][r]{-2}%
\psfrag{v02}[r][r]{0}%
\psfrag{v03}[r][r]{2}%
%
\resizebox{8cm}{!}{\includegraphics{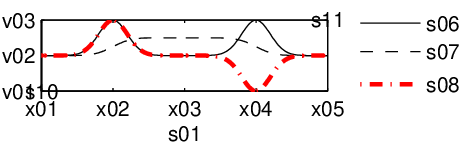}}%
\end{psfrags}%
%
\\	
%
\begin{psfrags}%
\psfragscanon%
%
\psfrag{s01}[t][t]{\fontsize{10}{15}\fontseries{m}\mathversion{normal}\fontshape{n}\selectfont \color[rgb]{0,0,0}\setlength{\tabcolsep}{0pt}\begin{tabular}{c}$x$\end{tabular}}%
\psfrag{s05}[l][l]{\fontsize{10}{15}\fontseries{m}\mathversion{normal}\fontshape{n}\selectfont \color[rgb]{0,0,0}$\partial_x \sigma$}%
\psfrag{s06}[l][l]{\fontsize{10}{15}\fontseries{m}\mathversion{normal}\fontshape{n}\selectfont \color[rgb]{0,0,0}$\rho$}%
\psfrag{s07}[l][l]{\fontsize{10}{15}\fontseries{m}\mathversion{normal}\fontshape{n}\selectfont \color[rgb]{0,0,0}$\sigma$}%
\psfrag{s08}[l][l]{\fontsize{10}{15}\fontseries{m}\mathversion{normal}\fontshape{n}\selectfont \color[rgb]{0,0,0}$\partial_x \sigma$}%
\psfrag{s10}[][]{\fontsize{10}{15}\fontseries{m}\mathversion{normal}\fontshape{n}\selectfont \color[rgb]{0,0,0}\setlength{\tabcolsep}{0pt}\begin{tabular}{c} \end{tabular}}%
\psfrag{s11}[][]{\fontsize{10}{15}\fontseries{m}\mathversion{normal}\fontshape{n}\selectfont \color[rgb]{0,0,0}\setlength{\tabcolsep}{0pt}\begin{tabular}{c} \end{tabular}}%
%
\fontsize{10}{15}\fontseries{m}\mathversion{normal}%
\fontshape{n}\selectfont%
%
\psfrag{x01}[t][t]{-2}%
\psfrag{x02}[t][t]{-1}%
\psfrag{x03}[t][t]{0}%
\psfrag{x04}[t][t]{1}%
\psfrag{x05}[t][t]{2}%
%
\psfrag{v01}[r][r]{-2}%
\psfrag{v02}[r][r]{0}%
\psfrag{v03}[r][r]{2}%
%
\resizebox{8cm}{!}{\includegraphics{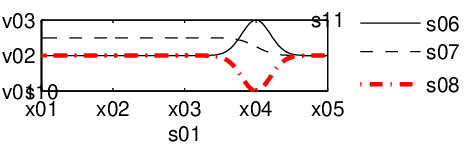}}%
\end{psfrags}%
%

	\caption{Stress and density profiles are shown
for two one-dimensional two-particle systems, each with two particles of unit mass at positions
$x=\pm 1$, and repelling each other (so with $d>2$ for our granular case).
In the top figure, both particles are flowing, while in the bottom figure the left particle is fixed
and the right one flowing.}
\label{fig:cg}
\end{figure} 	

Equation \eqref{eq:sd:18} differs from the results of \cite{Goldhirsch2010} by an additional term
that accounts for the stress created by the presence of the base.
The contribution to the stress from the interaction of two flow particles $i,j$ is spatially
distributed along the contact line from $\vec{r}_i$ to $\vec{r}_j$,
while the contribution from the interaction of particles $i$ with a fixed particle $k$ is distributed
along the line from $\vec{r}_i$ to $\vec{r}_k$, extending further beyond $\vec{r}_k$. 
We explain the situation as follows, see Fig.~\ref{fig:cg}.
Stress and density profiles are calculated using \eqref{eq:sd:4} and \eqref{eq:sd:18}
for two one-dimensional two-particle systems, each with two particles of unit mass at positions
$x=\pm 1$, repelling each other with a force $|f|=1$ and with $w=0.2$.
In the top figure, both particles belong to the flowing species,
so the density is distributed around the particles' center of mass and the stress along the contact line.
In the bottom figure, the left particle is a fixed base particle and the right particle is a
free flowing one, so density is distributed around the flowing particle's
center of mass and the stress along the line extending from the flowing particle to negative infinity.

The strength of this statistical method is that the spatial coarse graining satisfies the mass
and momentum balance equations exactly at any given time, irrespective of the choice of
the coarse graining function. 
Further details about the accuracy of the stress definition \eqref{eq:sd:18} are discussed in
\linebreak[4]\cite{WeinhartThorntonBokhoveLuding2011}.
The expression for the energy is also not treated in this publication, see \cite{Babic1997}.

\section{Mathematical background}\label{sec:math}

In this section, we briefly outline the existing  knowledge to develop a continuum shallow-layer theory for granular flow.

\subsection{Shallow-layer model}

Shal\-low-layer models have been shown to be an effective tool in modelling many geophysical mass flows.
Early ava\-lanche models were formulated by adding gravitational acceleration and Coulomb basal friction to
shal\-low-layer models \cite{GrigorianEglitIakimov1967,KulikovskiiEglit1973}.
Similar dry granular models have been derived using the long-wave approximation
\cite{SavageHutter1989,HutterSiegelSavageNohguchi1993,Iverson1997,GrayWielandHutter1999,WielandGrayHutter1999}
for shallow variations in the flow height and slope topography and included a Mohr-Coulomb rheology by the use of an earth pressure coefficient.
The key to these theories is to depth-integrate general three-dimen\-si\-o\-nal equations in the shallow direction,
resulting in a system of two-dimensional equations which still retains some information about variations in thickness.

Let $Oxyz$ be a coordinate system with the $x$-axis down\-slope and the $z$-axis normal to a channel
with mean slope $\theta$.
For simplicity, we further consider boundaries, flows, and external forcing to be (statistically) uniform in y.
The continuum macro-scale fields are thus independent of $y$,
while the DPM simulations remain three-dimensional and will be periodic in $y$. 
The free-surface and base location are $z=s(x,t)$ and $z=b(x)$, respectively.
The thickness of the flow is thus $h(x,t)=s(x,t)-b(x)$, and
the bulk density and velocity components are $\vec{\rho}$
and $\vec{u}=(u,v=0,w)^t$, respectively, as functions of $x,y,z$ and $t$.

The three-dimensional flow viewed as continuum is described by the mass and momentum balance equations
\eqref{eq:mass} and \eqref{eq:momentum}. 
At the top and bottom surface, kinetic boundary conditions are satisfied:
$D(z-s)/Dt = 0$ and $D(z-b)/Dt=0$ at their respective surfaces, and with material time derivative
\begin{equation}\nonumber
D(\cdot)/Dt=\partial(\cdot)/\partial t+u\partial(\cdot)/\partial x+w\partial(\cdot)/\partial z
\end{equation}
(since we assumed $v=0$).
Furthermore, the top surface is traction-free, while the traction at the basal surface is essentially Coulomb-like.
We decompose the traction $\vec{t}=\vec{t}_t+{t}_n\,\hat{\vec{n}}$ in tangential and  normal components,
with normal traction ${t}_n = -\hat{\vec{n}}\cdot(\sigma\,\hat{\vec{n}})$, where $\hat{\vec{n}}$ is the outward normal at the fixed base. 
The Coulomb Ansatz implies that $\vec{t}_t = -\mu |\vec{t}_n|\vec{u}/|\vec{u}|$ with friction factor $\mu>0$.
Note that $\mu$ generally can be a function of the local depth and the flow velocity.
Its determination is essential to find a closed system of shallow-layer equations.

We consider flows that are shallow, such that a typical aspect ratio $\epsilon$ between flow height and length,
normal and along\-slope velocity, or normal and along\-slope variations in basal topography, is small,
of order $\mathcal{O}(\epsilon)$.
Furthermore, the typical friction factor $\mu$ is small enough to satisfy $\mu=\mathcal{O}(\ep^\gamma)$
with $\gamma\in(0,1)$. 
We follow the derivation of the depth-averaged swallow layer equations for granular flow by \cite{GrayTaiNoelle2003}
without assuming that the flow is incompressible.
Instead we start the asymptotic analysis from the dimensionless form of the mass and momentum
conservation equations \eqref{eq:mass} and \eqref{eq:momentum},
assuming only that the density is independent of depth at leading order.
Density, velocity, and stress are depth averaged as follows
\begin{equation}\label{eq:depthavg}
\bar{()}= \frac{1}{h}\int_b^s () \ \rd z.
\end{equation}
In the end, one retains the normal stress ratio $K=\bar\sigma_{xx}/\bar\sigma_{zz}$,
the shape factor $\alpha=\overline{u^2}/\bar{u}^2$, and the friction $\mu$ as unknowns.
The goal is to investigate whether these unknowns
can be expressed as either constants or functions of the remaining shallow flow variables,
to leading order in $\mathcal{O}(\epsilon)$.
The latter variables are the flow thickness $h=h(x,t)$ and the depth-averaged velocity $\bar{u} = \bar{u}(x,t)$.
At leading order, the momentum equation normal to the base yields that the downward normal stress is lithostatic,
$\sigma_{zz}(z)= \bar{\rho} g\cos\theta (s-z) + \mathcal{O}(\epsilon)$. 
Depth-ave\-ra\-ging the remaining equations, while retaining only terms of order $\mathcal{O}(\epsilon\mu)$,
yields the dimensional depth-averaged shallow-layer equations, {\em cf.}, \cite{VremanAlTaraziKuipersVanSintAnnalandBokhove2007,BokhoveThornton2010},
\begin{subequations}\label{eq:swe}
\begin{equation}\label{eq:swe:a}
\pfrac{(\bar\rho h)}{t^{\star}} + \pfrac{}{x^\star} \left(\bar\rho h \overline{u} \right) = 0,
\end{equation}
\begin{equation}\label{eq:swe:b}
\pfrac{}{t^{\star}} \left(h \bar\rho \bar{u} \right) + 
\pfrac{}{x^\star} \left(h \bar\rho \alpha \overline{u}^2 + \frac{K}{2} g h^2 \bar\rho \cos \ca \right ) = gh\bar\rho S,
\end{equation}
with
\begin{equation}
S = \sin \ca - \mu \frac{\ubar}{\left|{\ubar}\right|}  \cos \ca - \pfrac{b}{x^\star}\cos\theta.
\end{equation}
\end{subequations}
To demarcate the dimensional time and spatial scales, we have used starred coordinates.
These scales differ from the ones used before in the particle dynamics and the dimensionless ones used later in the DPM simulations.
The shallow-layer equations \eqref{eq:swe} consist of the continuity equation \eqref{eq:swe:a}
and the downslope momentum equation \eqref{eq:swe:b}.
The system arises also via a straightforward control volume analysis of a column of granular material
viewed as continuum from base to the free surface,
using Reynold-stress averaging and a leading order closure with depth averages.

While the mean density $\bar{\rho}$ can be modelled as a system variable by considering the energy balance equation,
we will assume that it can be expressed as a function of height and velocity $\bar{\rho}(h,\bar u)$.
Thus, the closure to equations \eqref{eq:swe} is determined when we can find
the functions $\bar{\rho}(h,\bar u)$, $K(h,\bar u)$, $\alpha(h,\bar u)$, and $\mu(h,\bar u)$. 
In Section \ref{sec:depthaveraging}, we will analyze if and when DPM simulations can determine these functions.

\subsection{Granular friction laws for a rough basal surface}\label{sec:friction}

The friction coefficient, $\mu$, was originally \cite{HungrMorgenstern1984} taken to be a simple
Coulomb type $\mu=\tan \delta$, where $\delta$ is a fixed friction angle.
Note that in steady state for a flat base with $b=0$, the shallow-layer momentum equation \eqref{eq:swe:b} then yields $\mu=\tan\theta$.
Pure Coulomb friction implies that there is only one inclination, $\theta=\delta$,
at which steady flow of constant height and flow velocity exists. That turns out to be unrealistic. 
Three parameterizations for $\mu$ have been proposed in the literature.

Firstly, Forterre and Pouliquen \cite{ForterrePouliquen2003} found
steady flow in laboratory investigations for a range of inclinations concering flow over rough basal surfaces.
They measured the thickness $h_{stop}$ of stationary material, left behind when a flowing layer
was brought to rest, with the following fit
\begin{equation} \label{eq:hstop} 
\frac{h_{stop}(\theta)}{Ad} = \frac{\tan(\delta_2) - \tan(\theta)}{\tan(\theta) - \tan(\delta_1)},\quad \delta_1<\theta<\delta_2,
\end{equation}
where $\delta_1$ is the minimum angle required for flow,
$\delta_2$ the maximum angle at which steady uniform flow is possible, $d$ the particle diameter,
and $A$ a characteristic dimensionless length scale over which the friction varies.
Note that $h_{stop}$ diverges for $\theta=\delta_1$ and is zero for $\theta=\delta_2$. 
For $h>h_{stop}$, steady flow exists in which the Froude number,
the aspect ratio between flow speed and surface gravity-wave speed ($F=\bar{u}/\sqrt{g \cos\theta h}$),
is a linear function of the height,
\begin{equation} \label{eq:Pouliquen} 
F= \beta \frac{h}{h_{stop}(\theta)} - \gamma,\quad\mathrm{for}\quad\mathrm{for}\quad\delta_1<\theta<\delta_2,
\end{equation}
where $\beta$ and $\gamma$ are constants independent of chute inclination and particle size.
Provided one uses the steady state assumption $\mu=\tan\theta$ to hold (approximately) in the dynamic case as well,
one can combine it with \eqref{eq:hstop} and \eqref{eq:Pouliquen} to find an improved empirical friction law
\begin{equation} \label{eq:mu} 
\mu=\mu^\star(h,F) = \tan(\delta_1) + \frac{\tan(\delta_2)-\tan(\delta_1)}{\beta h/(Ad(F+\gamma))+1}.
\end{equation}
It is a closure for $\mu$ in terms of the flow variables, which has been shown its practical value.
Note that in the limit $\delta_1\to \delta_2=\delta$ the Coulomb model is recovered.

Secondly, in an earlier version \cite{Pouliquen1999a}, another, exponential fitting was proposed for $h_{stop}$,
as follows
\begin{equation} \label{eq:hstop_exp} 
\frac{h_{stop}'(\theta)}{A' d} = \ln \frac{\tan(\theta) - \tan(\delta_1')}{\tan(\delta_2') - \tan(\delta_1')},
\quad\mathrm{for}\quad \delta_1'<\theta<\delta_2'
\end{equation}
with the same limiting behaviour, and primes used to denote the difference in the fit. It yields the friction factor
\begin{equation}\label{1.5}
\mu=\mu'(h,F)= \tan \delta_1' + \left(\tan \delta_2' - \tan \delta_1'\right)e^{\left\{ \frac{-\beta' h}{A' d (F+\gamma')}\right\}}.
\end{equation}
Equation \eqref{eq:hstop} did, however, prove to be a better fit to experiments
and is computationally easier to evaluate.

Finally, an additional correction to \eqref{eq:Pouliquen} was made in \cite{Jenkins2006} to include
dependence on the inclination accounting for enduring particle contacts in very dissipative frictional flows, 
\begin{equation} \label{eq:PJ} 
F =  \beta_J \frac{h}{h_{stop}(\theta)} \frac{\tan^2(\theta)}{\tan^2(\delta_1)}-\gamma_J, 
\end{equation}
for which we can use any appropriate fit for $h_{stop}$.
It leads subsequently to a more complicated evaluation of the friction law for $\mu$. We omit further details.
We will compare our DPM simulations against these rules and fits for the rough basal surface,
and extend it to smoother surfaces in the upcoming sections.



\section{Simulation description}\label{sec:simulation}

\begin{figure}[t]
	\scalebox{.5}{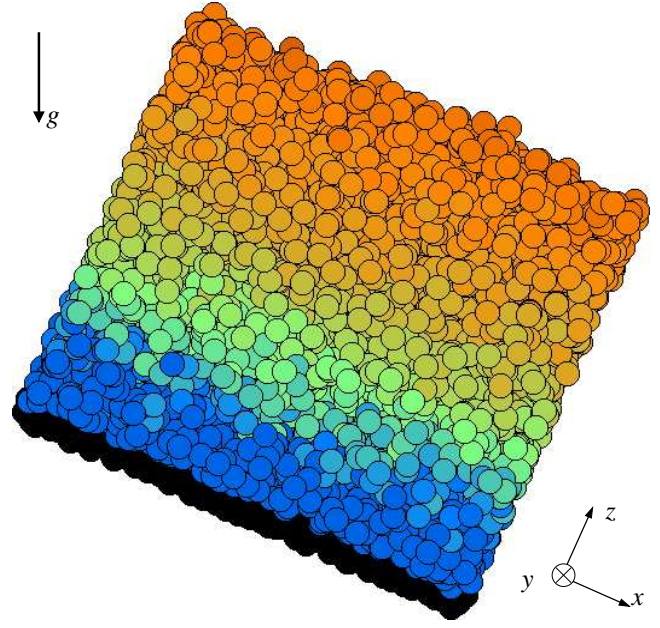}
	\caption{
DPM simulation for approximated height $H=17.5$, inclination $\theta=24^\circ$ and the diameter ratio of free and
fixed particles, $\lambda=1$, at time $t=2000$; gravity direction $g$ as indicated.
The domain is periodic in $x$- and $y$-directions.
In the $z$-direction, fixed particles (black) form a rough base while the surface is unconstrained.
Colours indicate speed, increasing from blue via green to orange.}
	\label{fig:chute}
\end{figure} 	

In this section, DPM simulations are used to simulate monodispersed granular flows.
Parameters have been nondimensionalised such that the flow particle diameter $d=1$, mass $m=1$ and the
magnitude of gravity $g=1$. The normal spring and damping constants are $k^n=2\cdot10^5$ and $\gamma^{\,n}=50$;
thus the contact duration is $t_c=0.005$ and the coefficient of restitution is $\epsilon=0.88$.
The tangential spring and damping constants are $k^{\,t}=(2/7) k^n$ and $\gamma^{\,t}=\gamma^{\,n}$,
such that the frequency of normal and tangential contact oscillation and the normal and tangential dissipation
are equal. The microscopic friction coefficient was taken to be $\mu^c=1/2$.
  
The interaction parameters are chosen as in \cite{SilbertErtasGrestHaleyLevinePlimpton2001}
to simulate glass particles of $0.1\,\mathrm{mm}$ size; this corresponds to a dimensional time scale of $\sqrt{d/g}=3.1\,\mathrm{ms}$ and
dimensional velocity scale $\sqrt{dg}=0.031\,\mathrm{ms}^{-1}$.
The above parameters are identical to the simulations of Silbert {\em et al}.,
except that dissipation in tangential direction, $\gamma^t$, was added to damp rotational degrees of freedom in arresting flow.
Adding of such tangential damping removes all vibrational energy for flows otherwise arrested.
The differences in the simulations with $\gamma^{\,t}=\gamma^{\,n}$ reported here are minor relative to the case with $\gamma^{\,t}=0$.
Silbert {\em et al.} also investigated the sensitivity of the results to the particle interaction parameters $t_c$,
$\epsilon$, the ratio $k^n/k^{\,t}$, and $\mu^c$;
they found that while the density of the bulk material is not sensitive to these interaction parameters,
the flow velocity increased with decreasing friction $\mu^c$.
Nonetheless, the qualitative behaviour of the velocity profiles did not change.

The chute is periodic and of size $20\times 10$ in the $x$- and $y$-directions and has a layer of fixed particles
as a base.
The bottom particles are monodispersed with (nondimensional) diameter $\lambda$.
Various basal roughnesses are investigated by taking $\lambda=0, 1/2$ to $2$ in turn,
with $\lambda=0$ as flat base.
This bottom particle layer is obtained by performing a simulation on a horizontal, smooth-bottom chute.
It is filled with a randomly distributed set of particles and we simulate until a static layer about 12 particles thick is produced.
Then the layers of particles at height $z\in[9.3,11]\lambda$ are selected, remaining particles deleted,
and the selected ones moved downwards by $11\lambda$.
The layer is thick enough to ensure that no flowing particles can fall through the base.
Their positions are fixed.

\begin{figure}[tbp]
%
\begin{psfrags}%
\psfragscanon%
%
\psfrag{s01}[t][t]{\fontsize{10}{15}\fontseries{m}\mathversion{normal}\fontshape{n}\selectfont \color[rgb]{0,0,0}\setlength{\tabcolsep}{0pt}\begin{tabular}{c}$t$\end{tabular}}%
\psfrag{s02}[b][b]{\fontsize{10}{15}\fontseries{m}\mathversion{normal}\fontshape{n}\selectfont \color[rgb]{0,0,0}\setlength{\tabcolsep}{0pt}\begin{tabular}{c}$E_{kin}/\langle{E}_{ela}\rangle$\end{tabular}}%
\psfrag{s05}[l][l]{\fontsize{7}{10.5}\fontseries{m}\mathversion{normal}\fontshape{n}\selectfont \color[rgb]{0,0,0}$\theta=20^\circ$}%
\psfrag{s06}[l][l]{\fontsize{7}{10.5}\fontseries{m}\mathversion{normal}\fontshape{n}\selectfont \color[rgb]{0,0,0}$\theta=60^\circ$}%
\psfrag{s07}[l][l]{\fontsize{7}{10.5}\fontseries{m}\mathversion{normal}\fontshape{n}\selectfont \color[rgb]{0,0,0}$\theta=50^\circ$}%
\psfrag{s08}[l][l]{\fontsize{7}{10.5}\fontseries{m}\mathversion{normal}\fontshape{n}\selectfont \color[rgb]{0,0,0}$\theta=40^\circ$}%
\psfrag{s09}[l][l]{\fontsize{7}{10.5}\fontseries{m}\mathversion{normal}\fontshape{n}\selectfont \color[rgb]{0,0,0}$\theta=30^\circ$}%
\psfrag{s10}[l][l]{\fontsize{7}{10.5}\fontseries{m}\mathversion{normal}\fontshape{n}\selectfont \color[rgb]{0,0,0}$\theta=29^\circ$}%
\psfrag{s11}[l][l]{\fontsize{7}{10.5}\fontseries{m}\mathversion{normal}\fontshape{n}\selectfont \color[rgb]{0,0,0}$\theta=28^\circ$}%
\psfrag{s12}[l][l]{\fontsize{7}{10.5}\fontseries{m}\mathversion{normal}\fontshape{n}\selectfont \color[rgb]{0,0,0}$\theta=27^\circ$}%
\psfrag{s13}[l][l]{\fontsize{7}{10.5}\fontseries{m}\mathversion{normal}\fontshape{n}\selectfont \color[rgb]{0,0,0}$\theta=26^\circ$}%
\psfrag{s14}[l][l]{\fontsize{7}{10.5}\fontseries{m}\mathversion{normal}\fontshape{n}\selectfont \color[rgb]{0,0,0}$\theta=25^\circ$}%
\psfrag{s15}[l][l]{\fontsize{7}{10.5}\fontseries{m}\mathversion{normal}\fontshape{n}\selectfont \color[rgb]{0,0,0}$\theta=24^\circ$}%
\psfrag{s16}[l][l]{\fontsize{7}{10.5}\fontseries{m}\mathversion{normal}\fontshape{n}\selectfont \color[rgb]{0,0,0}$\theta=23^\circ$}%
\psfrag{s17}[l][l]{\fontsize{7}{10.5}\fontseries{m}\mathversion{normal}\fontshape{n}\selectfont \color[rgb]{0,0,0}$\theta=22^\circ$}%
\psfrag{s18}[l][l]{\fontsize{7}{10.5}\fontseries{m}\mathversion{normal}\fontshape{n}\selectfont \color[rgb]{0,0,0}$\theta=21^\circ$}%
\psfrag{s19}[l][l]{\fontsize{7}{10.5}\fontseries{m}\mathversion{normal}\fontshape{n}\selectfont \color[rgb]{0,0,0}$\theta=20.5^\circ$}%
\psfrag{s20}[l][l]{\fontsize{7}{10.5}\fontseries{m}\mathversion{normal}\fontshape{n}\selectfont \color[rgb]{0,0,0}$\theta=20^\circ$}%
\psfrag{s22}[][]{\fontsize{10}{15}\fontseries{m}\mathversion{normal}\fontshape{n}\selectfont \color[rgb]{0,0,0}\setlength{\tabcolsep}{0pt}\begin{tabular}{c} \end{tabular}}%
\psfrag{s23}[][]{\fontsize{10}{15}\fontseries{m}\mathversion{normal}\fontshape{n}\selectfont \color[rgb]{0,0,0}\setlength{\tabcolsep}{0pt}\begin{tabular}{c} \end{tabular}}%
%
\fontsize{7}{10.5}\fontseries{m}\mathversion{normal}%
\fontshape{n}\selectfont%
%
\psfrag{x01}[t][t]{0}%
\psfrag{x02}[t][t]{500}%
\psfrag{x03}[t][t]{1000}%
\psfrag{x04}[t][t]{1500}%
\psfrag{x05}[t][t]{2000}%
%
\psfrag{v01}[r][r]{$10^{0}$}%
\psfrag{v02}[r][r]{$10^{1}$}%
\psfrag{v03}[r][r]{$10^{2}$}%
\psfrag{v04}[r][r]{$10^{3}$}%
\psfrag{v05}[r][r]{$10^{4}$}%
\psfrag{v06}[r][r]{$10^{5}$}%
%
\resizebox{8cm}{!}{\includegraphics{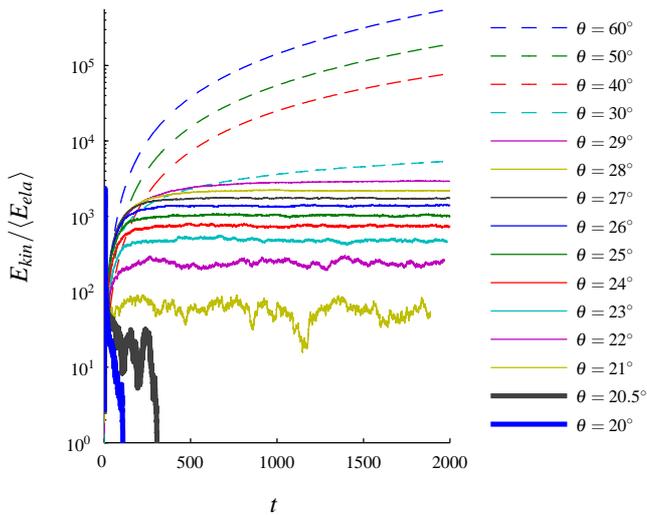}}%
\end{psfrags}%
%

	\caption{
Shown is the ratio of kinetic to mean elastic energy over time for $H=20$, basal roughness $\lambda=1$,
and varying chute angles $\theta$.
Flow stops for inclinations $\theta\leq 20.5^\circ$, remains steady for $21^\circ\leq\theta\leq 29^\circ$
and accelerates for $\theta\geq 30^\circ$ (dashed lines).}
	\label{fig:arresting}
\end{figure}

Defining a `filling height' $H$, an integer amount of about $200\,H$ particles
are inserted into the chute.
To insert a particle, a random location $(x,y,z)\in[0,20]\times[0,10]\times[0,I]$ is chosen, where $I=H$.
If the particle at this position overlaps other particles, the insertion is rejected,
and the insertion domain is enlarged by increasing $I$ to $ I+0.01$ to ensure that
there is enough space for all particles. 
The initial packing fraction is about $\rho/\rho_p=0.3$.
Thus the particles initially compact to an approximated height $H$, giving the chute enough kinetic energy to initialize flow.
Dimensionless time is integrated between $t\in[0,2000]$ to allow the system to reach a steady state.
A screen shot of the system in steady state is given in Fig.~\ref{fig:chute}.

To ensure that the size of the periodic box does not influence the result,
we compared density and velocity profiles of the flow at an angle $\theta=24^\circ$ and height $H=17.5$
for domain sizes $L_x=10$, $20$, $40$, $L_y=10$ and $L_x=10$, $20$, $40$, $L_y=L_x/2$,
and found no significant changes.

\section{Arrested, steady, and accelerating flow}\label{sec:rough}

From the experiments of Pouliquen \cite{Pouliquen1999a}, granular flow over a rough base is known to exist for a
range of heights and inclinations.
DPMs by \cite{SilbertErtasGrestHaleyLevinePlimpton2001} also showed that
steady flows arose for a range of flow heights and (depth-averaged) velocities or Froude numbers.
Their simulations did, however, provide relatively few data points near the boundary of
arrested and steady flow to allow a more adequate fit of the stopping height.
In this section, we therefore perform numerous DPMs at heights and angles near
the separatrix between the steady flow regime and the regime with static piles.
To study the full range of steady flow regimes,
simulations were performed for inclinations $\theta$ varying between $20^\circ$ and $60^\circ$
and approximated or `filling heights' $H=10,\ 20,\ 30$ and $40$.
In Section \ref{sec:smooth}, we will repeat (some of) these simulations
for varying base roughness.

\begin{figure}[tbp]
%
\begin{psfrags}%
\psfragscanon%
%
\psfrag{s01}[t][t]{\fontsize{10}{15}\fontseries{m}\mathversion{normal}\fontshape{n}\selectfont \color[rgb]{0,0,0}\setlength{\tabcolsep}{0pt}\begin{tabular}{c}$\theta$\end{tabular}}%
\psfrag{s02}[b][b]{\fontsize{10}{15}\fontseries{m}\mathversion{normal}\fontshape{n}\selectfont \color[rgb]{0,0,0}\setlength{\tabcolsep}{0pt}\begin{tabular}{c}$h$\end{tabular}}%
\psfrag{s03}[r][r]{\fontsize{10}{15}\fontseries{m}\mathversion{normal}\fontshape{n}\selectfont \color[rgb]{0,0,0}\setlength{\tabcolsep}{0pt}\begin{tabular}{r}$H=10$\end{tabular}}%
\psfrag{s04}[r][r]{\fontsize{10}{15}\fontseries{m}\mathversion{normal}\fontshape{n}\selectfont \color[rgb]{0,0,0}\setlength{\tabcolsep}{0pt}\begin{tabular}{r}$H=20$\end{tabular}}%
\psfrag{s05}[r][r]{\fontsize{10}{15}\fontseries{m}\mathversion{normal}\fontshape{n}\selectfont \color[rgb]{0,0,0}\setlength{\tabcolsep}{0pt}\begin{tabular}{r}$H=30$\end{tabular}}%
\psfrag{s06}[r][r]{\fontsize{10}{15}\fontseries{m}\mathversion{normal}\fontshape{n}\selectfont \color[rgb]{0,0,0}\setlength{\tabcolsep}{0pt}\begin{tabular}{r}$H=40$\end{tabular}}%
%
\fontsize{10}{15}\fontseries{m}\mathversion{normal}%
\fontshape{n}\selectfont%
%
\psfrag{x01}[t][t]{19}%
\psfrag{x02}[t][t]{20}%
\psfrag{x03}[t][t]{21}%
\psfrag{x04}[t][t]{22}%
\psfrag{x05}[t][t]{23}%
\psfrag{x06}[t][t]{24}%
\psfrag{x07}[t][t]{25}%
\psfrag{x08}[t][t]{26}%
\psfrag{x09}[t][t]{27}%
\psfrag{x10}[t][t]{28}%
\psfrag{x11}[t][t]{29}%
\psfrag{x12}[t][t]{30}%
%
\psfrag{v01}[r][r]{5}%
\psfrag{v02}[r][r]{10}%
\psfrag{v03}[r][r]{15}%
\psfrag{v04}[r][r]{20}%
\psfrag{v05}[r][r]{25}%
\psfrag{v06}[r][r]{30}%
\psfrag{v07}[r][r]{35}%
\psfrag{v08}[r][r]{40}%
\psfrag{v09}[r][r]{45}%
\psfrag{v10}[r][r]{50}%
%
\resizebox{8cm}{!}{\includegraphics{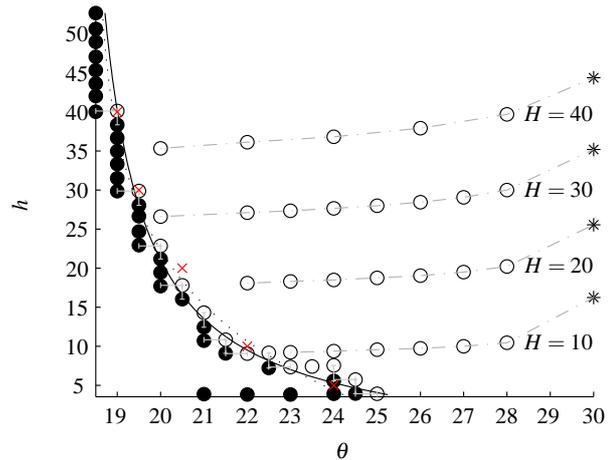}}%
\end{psfrags}%
%

	\caption{
Overview of DPMs for $\lambda=1$, with markers denoting the flow state at $t=2000$:
arrested $\bullet$, steady $\circ$, and accelerating $*$ flows.
Grey dash-dotted lines mark thickness $h$ for fixed $H$ from $H=10,20,30,40$ upward.
The demarcation line is fitted to $h_{stop}(\theta)$
in equation \eqref{eq:hstop} (solid line) and  $h_{stop}'(\theta)$ in \eqref{eq:hstop_exp} (dotted line).
Error bars mark intervals establishing the demarcation line.
Red crosses denote the demarcation between arrested and accelerating flow found in
\cite{SilbertErtasGrestHaleyLevinePlimpton2001}.}
	\label{fig:silbert_comparison}
\end{figure} 	

Whether a steady flow regime has been reached, is quantified here by assessing the time whereafter
the ratio of kinetic energy normalized by the mean elastic potential energy becomes time independent.
This is shown in Fig.~\ref{fig:arresting}, where we plot such an energy ratio for a rough base, constant height, and varying chute angle.
The elastic potential energy is averaged over $t\in[1000,2000]$ to minimize fluctuations after start-up,
but any interval larger than $100$ appears sufficient.
For chute angles at most $20.5^\circ$ the kinetic energy vanishes after a short time, thus the flow arrests;
for chute angles between $21^\circ-29^\circ$, a constant value is reached, indicating steady flow; and,
for inclinations above $29^\circ$ the energy keeps increasing: thus flow steadily accelerates.
If the energy ratio remained constant within $t\in[1800,2000]$, the flow was deemed steady,
otherwise the flow was deemed to be either accelerating or stopping. 

Unlike fluids, the free surface of granular flows, and thus the flow height, are not well defined.
In \cite{SilbertErtasGrestHaleyLevinePlimpton2001}, the height of the flow was estimated by $H$,
which is equivalent to assuming a constant packing fraction of $\rho/\rho_p=\pi/6$.
However, the exact height $h=s-b$ of the flow varies from the approximated height $H$ due to
compaction of the flow and $H$ is typically an overestimate.
In \cite{VremanAlTaraziKuipersVanSintAnnalandBokhove2007}, the surface of the flow was defined by the
time-average of the maximum vertical position of all flow particles.
One could also define the free surface of the flow as the height where the density vanishes.
The latter two methods, however, have the disadvantage that saltating particles
can lead to slightly overestimated flow heights. 

Instead, we will define the height via the downward normal stress. For steady uniform flows the downward normal stress is lithostatic,
\emph{i.e.}, balances the gravitational weight, such that
\begin{equation}
	\sigma_{zz}(z)=\int_z^\infty \rho(z')g\cos\theta \,dz'.\label{hydrostatic}
\end{equation}
This is a direct consequence of the momentum balance equations.
Thus, $\sigma_{zz}(z)$ has to decrease monotonically;
the base and free surface are the heights at which $\sigma_{zz}(z)$ reaches its maximum and minimum value, respectively.
However, in order to avoid effects of coarse graining or single particles near the boundary,
we cut off the stress $\sigma_{zz}(z)$ on either boundary by defining threshold heights
\begin{subequations}
\begin{gather}
	z_1=\min\{z:~\sigma_{zz}<(1-\kappa) \max_{z\in\mathbb{R}}\sigma_{zz}\}\quad\textrm{and}\quad\\
	z_2=\max\{z:~\sigma_{zz}>\kappa \max_{z\in\mathbb{R}}\sigma_{zz}\}
\end{gather}
with $\kappa=1\%$.
We subsequently linearly extrapolate the stress profile in the interval $(z_1,z_2)$ to define the base $b$ and
surface height $s$ as the height at which the linear extrapoation reaches the maximum and minimum values of $\sigma_{zz}$, respectively,
\begin{equation}
	b=z_1-\frac{\kappa}{1-2\kappa}(z_2-z_1),\ 
	s=z_2+\frac{\kappa}{1-2\kappa}(z_2-z_1).
\end{equation}
\end{subequations}
The variable most sensitive to these height choices is $\bar\rho$. 
It shows well-defined functional behaviour for our definition of height, shown later.
This is not the case if we define height by the density or the method in \cite{VremanAlTaraziKuipersVanSintAnnalandBokhove2007}.
The threshold $\kappa=1\%$ was chosen because the results in Fig.~\ref{fig:Nu} were relatively insensitive to the choice of
$\kappa$ at or above $1\%$.

To determine the demarcation line $h_s(\theta;\lambda)$ between arrested and steady flow with good accuracy,
we performed a set of simulations with initial conditions determined by the following algorithm.
Starting from an initial `filling height' $H=4$ and inclination $\theta=21^\circ$,
the angle was increased in steps of $1^\circ$ until eventually a flowing state was reached.
From this initial flowing state, the height was increased by $2$ particle heights, if the flow arrested, 
or else the angle decreased by $1/2^\circ$, assuming the curve is monotonically decreasing.
Flow was defined to be arrested when the energy ratio $E_{kin}/E_{ela}$ fell below $10^{-5}$ within
$500$ time units, otherwise the flow was classified as flowing.
In simulations in which such arrested flows were continued after $t=500$,
a further decrease of kinetic energy was observed, thus validating the approach.
This procedure yields intervals of the inclination angle for each height and, vice versa,
height intervals for each angle, between which the demarcation line lies.
The values presented in \cite{SilbertErtasGrestHaleyLevinePlimpton2001} deviate at most $0.5^\circ$ from our observations,
perhaps due to the preparation of the chute bottom, or the slightly different dissipation used.
A demarcating curve between steady and arrested flow was fitted to equations \eqref{eq:hstop} and \eqref{eq:hstop_exp}
by minimizing the horizontal, respectively vertical, distance of the fit to these intervals,
see Fig.~\ref{fig:silbert_comparison}.
Fitting $h_{stop}(\theta)$ yields better results than $h_{stop}'(\theta)$ for all roughnesses
and only the fit \eqref{eq:hstop} will be used hereafter.
Similar fits will be made in Section \ref{sec:smooth} for varying basal roughness.
It leads us to a study of the depth profiles for steady state flow in the following section.

\section{Statistics for uniform steady flow}\label{sec:w}

To obtain detailed information about steady flows, we use the statistics defined in Sect.~\ref{sec:statistics}. 
Since the flows of interest are steady and uniform in $x$ and $y$, density, velocity and stress will be averaged over $x$, $y$ and $t$. 
The resulting depth profiles will depend strongly on the coarse-graining width $w$, which needs to be carefully selected. 
Representative depth profiles for particular heights, inclinations and basal roughnesses will also be analyzed. 

Depth profiles for steady uniform flow are averaged using a coarse graining width $w$ over $x\in(0,20]$,
$y\in(0,10]$ and $t\in[2000,2000+T]$. 
The profile of a variable $\chi$ is thus defined as 
\begin{equation} \langle\chi\rangle_w^T (z) = \frac{1}{200T}\int_{2000}^{2000+T}\int_{0}^{10}\int_{0}^{20} \chi_w(t,x,y,z) \,dx\,dy\,dt,
\end{equation}
with $\chi_w$ in turn the macroscopic field(s)
of density, momentum and stress, as defined in \eqref{eq:sd:2}, \eqref{eq:sd:5} and \eqref{eq:sd:18}. 
We average in time with time snapshots taken every $t_c/2$ units. 

\begin{figure}[t]
%
\begin{psfrags}%
\psfragscanon%
%
\psfrag{s01}[l][l]{\fontsize{10}{15}\fontseries{m}\mathversion{normal}\fontshape{n}\selectfont \color[rgb]{0,0,0}\setlength{\tabcolsep}{0pt}\begin{tabular}{l}1\end{tabular}}%
\psfrag{s02}[l][l]{\fontsize{10}{15}\fontseries{m}\mathversion{normal}\fontshape{n}\selectfont \color[rgb]{0,0,0}\setlength{\tabcolsep}{0pt}\begin{tabular}{l}1\end{tabular}}%
\psfrag{s03}[t][t]{\fontsize{10}{15}\fontseries{m}\mathversion{normal}\fontshape{n}\selectfont \color[rgb]{0,0,0}\setlength{\tabcolsep}{0pt}\begin{tabular}{c}$T$\end{tabular}}%
\psfrag{s04}[b][b]{\fontsize{10}{15}\fontseries{m}\mathversion{normal}\fontshape{n}\selectfont \color[rgb]{0,0,0}\setlength{\tabcolsep}{0pt}\begin{tabular}{c}$r$\end{tabular}}%
%
\fontsize{10}{15}\fontseries{m}\mathversion{normal}%
\fontshape{n}\selectfont%
%
\psfrag{x01}[t][t]{$10^{0}$}%
\psfrag{x02}[t][t]{$10^{1}$}%
\psfrag{x03}[t][t]{$10^{2}$}%
%
\psfrag{v01}[r][r]{$10^{-2}$}%
\psfrag{v02}[r][r]{$10^{-1}$}%
%
\resizebox{8cm}{!}{\includegraphics{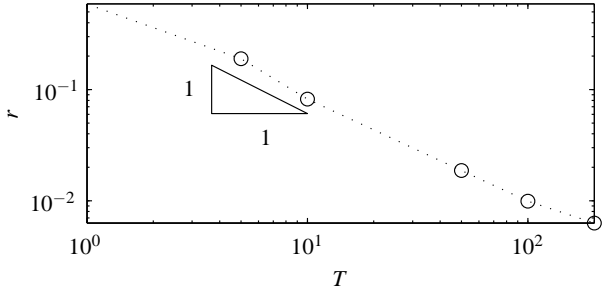}}%
\end{psfrags}%
%

	\caption{Depth-averaged norm of the momentum rate of change, $r=\int_b^s |\partial_t(\rho\vec{u})|_2 \,dz$,
with $\partial_t(\rho\vec{u})$ determined by \eqref{eq:st} for varying time averaging interval $T$.
Steady flow at height $H=20$ and inclination $\theta=24^\circ$ was used.
Temporal fluctuations decrease inversely proportional to the length of the time averaging interval.}
	\label{fig:t_dependence_momeq2}
\end{figure}  

To determine an appropriate time averaging interval $T$,
we calculate the rate of change in momentum from the density, velocity and stress fields by
\begin{equation}\label{eq:st}
\pfrac{(\rho\vec{u})}{t}=\nabla\cdot \vec{\sigma} - \rho \vec{g} - \vec{u}\cdot\nabla(\rho \vec{u}).
\end{equation}
For steady flow, the temporal variations in mass and momentum should approach zero when averaged over a long enough
time interval $T$. 
This is shown in Fig.~\ref{fig:t_dependence_momeq2}, where we plot the depth-averaged norm of the momentum rate of
change for varying time averaging interval.
For $T\geq100$, the temporal fluctuations decrease to less than $2\%$ of the largest term, $\bar\rho g$, in the momentum equation.
In the remainder, we choose $T=100$ as the averaging interval.

\begin{figure}[tbp]
{
\begin{psfrags}%
\psfragscanon%
%
\psfrag{s01}[t][t]{\fontsize{10}{15}\fontseries{m}\mathversion{normal}\fontshape{n}\selectfont \color[rgb]{0,0,0}\setlength{\tabcolsep}{0pt}\begin{tabular}{c}$z-b$\end{tabular}}%
\psfrag{s02}[b][b]{\fontsize{10}{15}\fontseries{m}\mathversion{normal}\fontshape{n}\selectfont \color[rgb]{0,0,0}\setlength{\tabcolsep}{0pt}\begin{tabular}{c}$\rho/\rho_p$\end{tabular}}%
\psfrag{s05}[l][l]{\fontsize{10}{15}\fontseries{m}\mathversion{normal}\fontshape{n}\selectfont \color[rgb]{0,0,0}$w=1$}%
\psfrag{s06}[l][l]{\fontsize{10}{15}\fontseries{m}\mathversion{normal}\fontshape{n}\selectfont \color[rgb]{0,0,0}$w=0.1$}%
\psfrag{s07}[l][l]{\fontsize{10}{15}\fontseries{m}\mathversion{normal}\fontshape{n}\selectfont \color[rgb]{0,0,0}$w=0.25$}%
\psfrag{s08}[l][l]{\fontsize{10}{15}\fontseries{m}\mathversion{normal}\fontshape{n}\selectfont \color[rgb]{0,0,0}$w=0.5$}%
\psfrag{s09}[l][l]{\fontsize{10}{15}\fontseries{m}\mathversion{normal}\fontshape{n}\selectfont \color[rgb]{0,0,0}$w=1$}%
\psfrag{s11}[][]{\fontsize{10}{15}\fontseries{m}\mathversion{normal}\fontshape{n}\selectfont \color[rgb]{0,0,0}\setlength{\tabcolsep}{0pt}\begin{tabular}{c} \end{tabular}}%
\psfrag{s12}[][]{\fontsize{10}{15}\fontseries{m}\mathversion{normal}\fontshape{n}\selectfont \color[rgb]{0,0,0}\setlength{\tabcolsep}{0pt}\begin{tabular}{c} \end{tabular}}%
%
\fontsize{10}{15}\fontseries{m}\mathversion{normal}%
\fontshape{n}\selectfont%
%
\psfrag{x01}[t][t]{0}%
\psfrag{x02}[t][t]{5}%
\psfrag{x03}[t][t]{10}%
\psfrag{x04}[t][t]{15}%
\psfrag{x05}[t][t]{20}%
\psfrag{x06}[t][t]{25}%
\psfrag{x07}[t][t]{30}%
%
\psfrag{v01}[r][r]{0.45}%
\psfrag{v02}[r][r]{0.5}%
\psfrag{v03}[r][r]{0.55}%
\psfrag{v04}[r][r]{0.6}%
\psfrag{v05}[r][r]{0.65}%
\psfrag{v06}[r][r]{0.7}%
%
\resizebox{8cm}{!}{\includegraphics{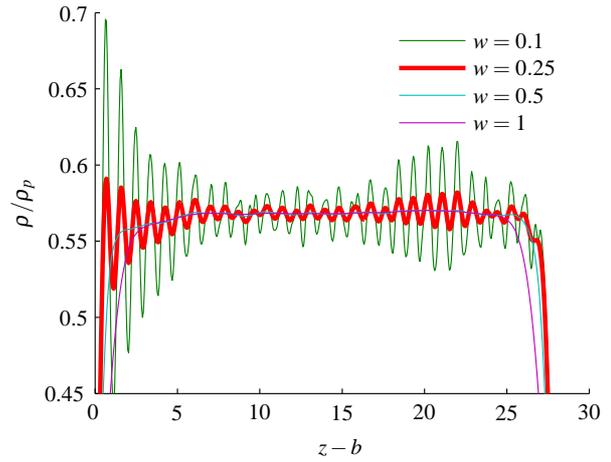}}%
\end{psfrags}%
%
}
\caption{Particle volume fraction $\rho/\rho_p$ for $H=30$, $\theta=24^\circ$, and $\lambda=1$ for varying coarse
graining widths $w$. Microscopic layering effects are visible for $w<0.5$. The density is constant in the bulk
and decreases slightly in the basal layer.}
\label{fig:w_dependence_nu}
\end{figure}

\begin{figure}[tbp]
{
%
\begin{psfrags}%
\psfragscanon%
%
\psfrag{s01}[t][t]{\fontsize{10}{15}\fontseries{m}\mathversion{normal}\fontshape{n}\selectfont \color[rgb]{0,0,0}\setlength{\tabcolsep}{0pt}\begin{tabular}{c}$z-b$\end{tabular}}%
\psfrag{s05}[l][l]{\fontsize{10}{15}\fontseries{m}\mathversion{normal}\fontshape{n}\selectfont \color[rgb]{0,0,0}\setlength{\tabcolsep}{0pt}\begin{tabular}{l}$\leftarrow\sigma_{xx}$\end{tabular}}%
\psfrag{s06}[r][r]{\fontsize{10}{15}\fontseries{m}\mathversion{normal}\fontshape{n}\selectfont \color[rgb]{0,0,0}\setlength{\tabcolsep}{0pt}\begin{tabular}{r}$\sigma_{zz}\rightarrow$\end{tabular}}%
\psfrag{s07}[r][r]{\fontsize{10}{15}\fontseries{m}\mathversion{normal}\fontshape{n}\selectfont \color[rgb]{0,0,0}\setlength{\tabcolsep}{0pt}\begin{tabular}{r}$\sigma_{yy}\rightarrow$\end{tabular}}%
\psfrag{s08}[r][r]{\fontsize{10}{15}\fontseries{m}\mathversion{normal}\fontshape{n}\selectfont \color[rgb]{0,0,0}\setlength{\tabcolsep}{0pt}\begin{tabular}{r}$-\sigma_{xz}\rightarrow$\end{tabular}}%
%
\fontsize{10}{15}\fontseries{m}\mathversion{normal}%
\fontshape{n}\selectfont%
%
\psfrag{x01}[t][t]{0}%
\psfrag{x02}[t][t]{5}%
\psfrag{x03}[t][t]{10}%
\psfrag{x04}[t][t]{15}%
\psfrag{x05}[t][t]{20}%
\psfrag{x06}[t][t]{25}%
%
\psfrag{v01}[r][r]{5}%
\psfrag{v02}[r][r]{10}%
\psfrag{v03}[r][r]{15}%
\psfrag{v04}[r][r]{20}%
\psfrag{v05}[r][r]{25}%
%
\resizebox{8cm}{!}{\includegraphics{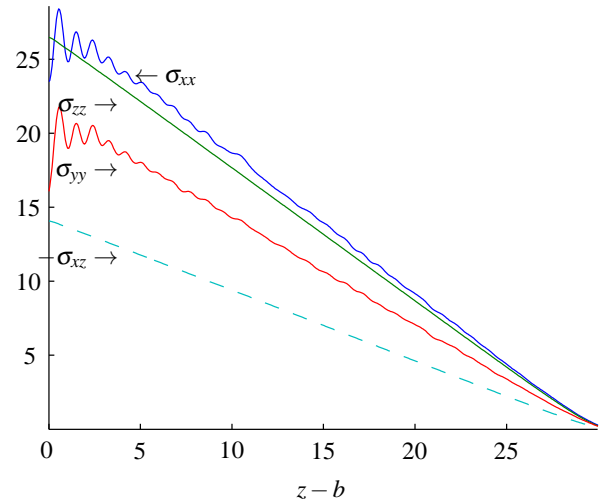}}%
\end{psfrags}%
%
}
\caption{Normal and shear stresses for $H=30$, $\theta=28^\circ$, and $\lambda=1$.
Shear $\sigma_{xz}$ and downward normal stress $\sigma_{zz}$ are balanced by gravitational forces,
see equation \eqref{hydrostatic}. The other normal stresses show anisotropic behaviour.}
\label{fig:NormalStresses}
\end{figure}

The effect of varying coarse-graining width $w$ is shown in Fig.~\ref{fig:w_dependence_nu},
which shows the $z$-profile of particle volume fraction $\rho/\rho_p$, where $\rho_p$ is the particle density.
For small $w$ we observe strong oscillations of about $0.9$ particle diameters width, particularly at the base.
The microscopic oscillations are increasingly smoothed out and finally vanish as we approach $w=0.5$.
For larger $w$, such as $w\ge 1$, the macroscopic gradients at the base and surface are smoothed out,
an unwanted effect of the coarse-graining. The same behaviour is observed in the stress and velocity fields.
Smoothing over the microscopic effects makes it impossible to observe microscopic layering in the density,
which we still wish to identify in our averaged fields. Hence, we choose $w=0.25$ as the coarse-graining width,
such that layering effects remain visible along with macroscopic gradients.

\begin{figure}[t]
{
\begin{psfrags}%
\psfragscanon%
%
\psfrag{s01}[t][t]{\fontsize{10}{15}\fontseries{m}\mathversion{normal}\fontshape{n}\selectfont \color[rgb]{0,0,0}\setlength{\tabcolsep}{0pt}\begin{tabular}{c}$(z-b)/h$\end{tabular}}%
\psfrag{s02}[b][b]{\fontsize{10}{15}\fontseries{m}\mathversion{normal}\fontshape{n}\selectfont \color[rgb]{0,0,0}\setlength{\tabcolsep}{0pt}\begin{tabular}{c}$u/\overline{u}$\end{tabular}}%
\psfrag{s05}[l][l]{\fontsize{8}{12}\fontseries{m}\mathversion{normal}\fontshape{n}\selectfont \color[rgb]{0,0,0}Bagnold}%
\psfrag{s06}[l][l]{\fontsize{8}{12}\fontseries{m}\mathversion{normal}\fontshape{n}\selectfont \color[rgb]{0,0,0}$\lambda=0,\theta=24^\circ$}%
\psfrag{s07}[l][l]{\fontsize{8}{12}\fontseries{m}\mathversion{normal}\fontshape{n}\selectfont \color[rgb]{0,0,0}$\lambda=1/2,\theta=22^\circ$}%
\psfrag{s08}[l][l]{\fontsize{8}{12}\fontseries{m}\mathversion{normal}\fontshape{n}\selectfont \color[rgb]{0,0,0}$\lambda=1/2,\theta=26^\circ$}%
\psfrag{s09}[l][l]{\fontsize{8}{12}\fontseries{m}\mathversion{normal}\fontshape{n}\selectfont \color[rgb]{0,0,0}$\lambda=2/3,\theta=24^\circ$}%
\psfrag{s10}[l][l]{\fontsize{8}{12}\fontseries{m}\mathversion{normal}\fontshape{n}\selectfont \color[rgb]{0,0,0}$\lambda=5/6,\theta=24^\circ$}%
\psfrag{s11}[l][l]{\fontsize{8}{12}\fontseries{m}\mathversion{normal}\fontshape{n}\selectfont \color[rgb]{0,0,0}$\lambda=1,\theta=24^\circ$}%
\psfrag{s12}[l][l]{\fontsize{8}{12}\fontseries{m}\mathversion{normal}\fontshape{n}\selectfont \color[rgb]{0,0,0}$\lambda=2,\theta=24^\circ$}%
\psfrag{s13}[l][l]{\fontsize{8}{12}\fontseries{m}\mathversion{normal}\fontshape{n}\selectfont \color[rgb]{0,0,0}Bagnold}%
\psfrag{s15}[][]{\fontsize{10}{15}\fontseries{m}\mathversion{normal}\fontshape{n}\selectfont \color[rgb]{0,0,0}\setlength{\tabcolsep}{0pt}\begin{tabular}{c} \end{tabular}}%
\psfrag{s16}[][]{\fontsize{10}{15}\fontseries{m}\mathversion{normal}\fontshape{n}\selectfont \color[rgb]{0,0,0}\setlength{\tabcolsep}{0pt}\begin{tabular}{c} \end{tabular}}%
%
\fontsize{8}{12}\fontseries{m}\mathversion{normal}%
\fontshape{n}\selectfont%
%
\psfrag{x01}[t][t]{0}%
\psfrag{x02}[t][t]{0.1}%
\psfrag{x03}[t][t]{0.2}%
\psfrag{x04}[t][t]{0.3}%
\psfrag{x05}[t][t]{0.4}%
\psfrag{x06}[t][t]{0.5}%
\psfrag{x07}[t][t]{0.6}%
\psfrag{x08}[t][t]{0.7}%
\psfrag{x09}[t][t]{0.8}%
\psfrag{x10}[t][t]{0.9}%
\psfrag{x11}[t][t]{1}%
%
\psfrag{v01}[r][r]{0}%
\psfrag{v02}[r][r]{0.2}%
\psfrag{v03}[r][r]{0.4}%
\psfrag{v04}[r][r]{0.6}%
\psfrag{v05}[r][r]{0.8}%
\psfrag{v06}[r][r]{1}%
\psfrag{v07}[r][r]{1.2}%
\psfrag{v08}[r][r]{1.4}%
\psfrag{v09}[r][r]{1.6}%
%
\resizebox{8cm}{!}{\includegraphics{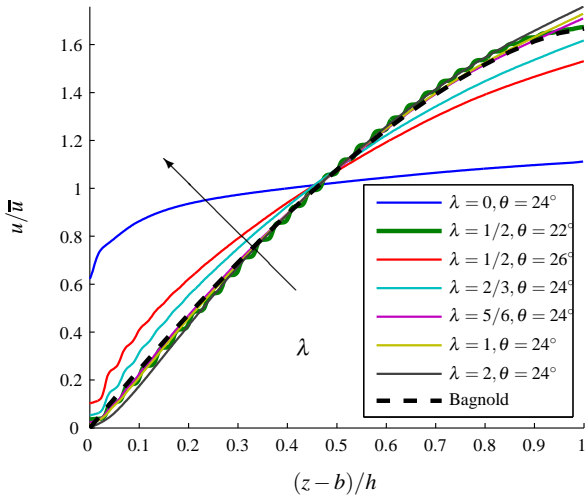}}%
\end{psfrags}%
%
}
\setlength{\unitlength}{\textwidth}
\begin{picture}(0,0)
\put(-0.25,0.15){\vector(-1,1){.1}}
\put(-0.25,0.1){$\lambda$}\end{picture}
\caption{
Flow velocity profile of thick flow for $H=30$, $\theta=24^\circ$,
$\lambda=0,2/3,5/6,1,2$ and for $H=30$, $\theta=22^\circ, 26^\circ$, $\lambda=1/2$.
For a rough base with $\lambda\ge 5/6$, we see a Bagnold velocity profile (dashed line), except near the surface.
For a smooth bases with $\lambda\leq2/3$, the profile becomes more convex.
For $\lambda=1/2$, $\theta<24^\circ$, the flow velocity shows layering while still observing the Bagnold profile.
For $\lambda=0$, a considerable slip velocity is observed.
For $\lambda=2$, the basal shear is small due to flow particles trapped between basal particles
so that the definition of the base $b(x)$ is rather fuzzy.
}
\label{fig:velocityX_lambda}
\end{figure}

The microscopic oscillations at the base indicate a strong layering effect of particles near the boundary,
despite the rough bottom surface. The macroscopic density throughout the flow is almost constant in the bulk and
decreases slightly at the base. An approximately constant density profile is a feature of all steady flows
and is an important assumption of depth-averaging. 

Non-zero stress components are plotted in Fig.~\ref{fig:NormalStresses}.
We observe that the stress components are nearly symmetric.
Shear stresses $\sigma_{yx}$ and $\sigma_{yz}$ are zero since
the flow velocity in $y$-direction vanishes. 
For steady flow, the downward normal stress $\sigma_{zz}(z)$ is lithostatic and satisfies equation \eqref{hydrostatic}
with a maximum error of $0.5\%$.
Since the density is nearly constant, we obtain a linear stress profile, another assumption of depth-averaged theory. 
Applying the momentum balance \eqref{eq:momentum} to steady uniform flow further yields that the shear stress satisfies
$\sigma_{xz}=\int_z^\infty \rho(z')g\sin\theta\,dz'.$
Thus, the macro-scale friction $\mu$ satisfies $\mu=\sigma_{xz}/\sigma_{zz}=-g_x/g_z=\tan\theta$.
This relation is locally satisfied for all steady flow cases to an accuracy of $|\theta-\tan^{-1}(\mu)|<0.4^\circ$.  
The remaining normal stress components, $\sigma_{xx}$ and $\sigma_{yy}$, are not constrained by this mass balance.
We thus see in Fig.~\ref{fig:NormalStresses}
significant anisotropy in the amplitude of the normal stresses, in particular in $\sigma_{yy}$,
showing that the confining stress is largest in the flow direction, except for very small inclinations.
It is always weakest in the lateral or $y$--direction with fluctuations at the base that are in phase with the
fluctuations of the density. Generally, the anisotropy increases with higher inclinations and smoother bases,
as will be analyzed in future work.

\begin{figure}[tbp]
%
\begin{psfrags}%
\psfragscanon%
%
\psfrag{s01}[t][t]{\fontsize{10}{15}\fontseries{m}\mathversion{normal}\fontshape{n}\selectfont \color[rgb]{0,0,0}\setlength{\tabcolsep}{0pt}\begin{tabular}{c}$\theta$\end{tabular}}%
\psfrag{s02}[b][b]{\fontsize{10}{15}\fontseries{m}\mathversion{normal}\fontshape{n}\selectfont \color[rgb]{0,0,0}\setlength{\tabcolsep}{0pt}\begin{tabular}{c}$h$\end{tabular}}%
\psfrag{s05}[l][l]{\fontsize{10}{15}\fontseries{m}\mathversion{normal}\fontshape{n}\selectfont \color[rgb]{0,0,0}$\lambda=2$}%
\psfrag{s06}[l][l]{\fontsize{10}{15}\fontseries{m}\mathversion{normal}\fontshape{n}\selectfont \color[rgb]{0,0,0}$\lambda=1/2$}%
\psfrag{s07}[l][l]{\fontsize{10}{15}\fontseries{m}\mathversion{normal}\fontshape{n}\selectfont \color[rgb]{0,0,0}$\lambda=2/3$}%
\psfrag{s08}[l][l]{\fontsize{10}{15}\fontseries{m}\mathversion{normal}\fontshape{n}\selectfont \color[rgb]{0,0,0}$\lambda=5/6$}%
\psfrag{s09}[l][l]{\fontsize{10}{15}\fontseries{m}\mathversion{normal}\fontshape{n}\selectfont \color[rgb]{0,0,0}$\lambda=1$}%
\psfrag{s10}[l][l]{\fontsize{10}{15}\fontseries{m}\mathversion{normal}\fontshape{n}\selectfont \color[rgb]{0,0,0}$\lambda=2$}%
\psfrag{s12}[][]{\fontsize{10}{15}\fontseries{m}\mathversion{normal}\fontshape{n}\selectfont \color[rgb]{0,0,0}\setlength{\tabcolsep}{0pt}\begin{tabular}{c} \end{tabular}}%
\psfrag{s13}[][]{\fontsize{10}{15}\fontseries{m}\mathversion{normal}\fontshape{n}\selectfont \color[rgb]{0,0,0}\setlength{\tabcolsep}{0pt}\begin{tabular}{c} \end{tabular}}%
%
\fontsize{10}{15}\fontseries{m}\mathversion{normal}%
\fontshape{n}\selectfont%
%
\psfrag{x01}[t][t]{19}%
\psfrag{x02}[t][t]{20}%
\psfrag{x03}[t][t]{21}%
\psfrag{x04}[t][t]{22}%
\psfrag{x05}[t][t]{23}%
\psfrag{x06}[t][t]{24}%
\psfrag{x07}[t][t]{25}%
%
\psfrag{v01}[r][r]{5}%
\psfrag{v02}[r][r]{10}%
\psfrag{v03}[r][r]{15}%
\psfrag{v04}[r][r]{20}%
\psfrag{v05}[r][r]{25}%
\psfrag{v06}[r][r]{30}%
\psfrag{v07}[r][r]{35}%
\psfrag{v08}[r][r]{40}%
\psfrag{v09}[r][r]{45}%
\psfrag{v10}[r][r]{50}%
%
\resizebox{8cm}{!}{\includegraphics{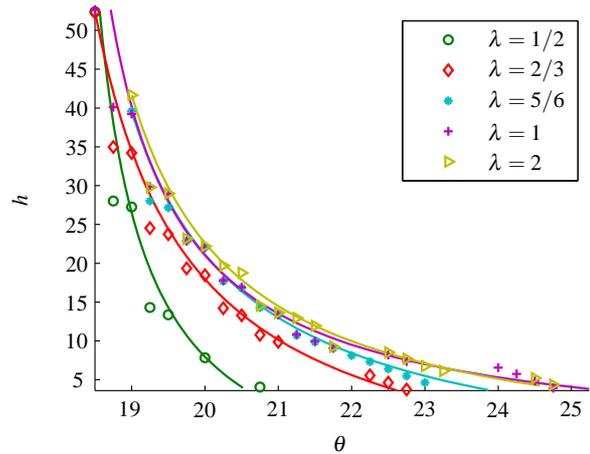}}%
\end{psfrags}%
%

\caption{Demarcation line $h_s(\theta;\lambda)$ for varying basal roughness.
Markers denote the midpoint of the intervals around which the curve was fitted.
Steady flow is observed at smaller inclinations for smoother bases.
While the smaller angle $\delta_{1,\lambda}$ varies only slightly,
the larger angle $\delta_{2,\lambda}$ decreases rapidly with the smoothness.
For $\lambda=0$, the demarcation line is vertical at $\theta=12.5^\circ$.}
\label{fig:hstop_smooth}
\end{figure}

\section{Transition from rough to smooth base}\label{sec:smooth}

Next, we study the effect of smoother bases on the range of steady flows by decreasing the diameter
$\lambda$ of the base particles, with the limiting case of a flat bottom wall for $\lambda=0$.
Such an extensive numerical study of the effects of changing bottom roughness appears to be novel.
To that effect, the DPM simulations from Section \ref{sec:rough} were extended such that results for basal roughnesses
$\lambda=0$, $1/2$, $2/3$, $5/6$, $1$, and $2$ can be compared. 
Before we show the $h_{stop}$--curves for these simulations, we investigate the extent to which changes in basal roughness
lead to more complex density and velocity profiles. 

We summarize the density profiles seen without explicitly showing the results.
For decreasing basal roughness $\lambda$, we observe that the microscopic oscillations and the dip in density at the base increase, 
while the bulk density remains constant.
For $\lambda=1/2$ and $\lambda=2/3$ and small inclinations,
we see steady flow that is strongly layered throughout the flow.
In contrast, for $\lambda=2$ there is a low flow density in the basal region,
since some of the free particles are small enough to sink a little into the base,
forming a mixed layer of fixed and free particles.

Velocity profiles for $H=30$ and $\theta=24^\circ$ are shown for varying basal roughness in Fig.~\ref{fig:velocityX_lambda}.
For $\lambda=1$, we observe the Bagnold profile \cite{Bagnold1954} for thick collisional flows,
differing only at the surface. For very thin flows at $H=10$ or inclinations near the arresting flow regime,
the profile differs strongly from the Bagnold profile and becomes linear. For smoother bases, the flow velocity increases,
and the profile becomes more concave. Weak to stronger slip velocities are observed for $\lambda<2/3$.
For $\lambda=0$, thicker flows have constant velocity throughout the depth, almost corresponding to plug flow. 

A family of demarcation curves $h_{stop}(\theta;\lambda)$ between steady and arrested flow is shown
in Fig.~\ref{fig:hstop_smooth}.
The curve fits are based on 
\begin{equation} \label{eq:hstop_thomas} 
h_{stop}(\theta;\lambda) = A_\lambda d \frac{\tan(\delta_{2,\lambda}) - \tan(\theta)}{\tan(\theta) - \tan(\delta_{1,\lambda})},
\quad \delta_{1,\lambda}<\theta<\delta_{2,\lambda},
\end{equation}
in which the dependencies on $\lambda$ are explicitly denoted.
The fitting parameters $\delta_{1,\lambda}$, $\delta_{2,\lambda}$, $A_\lambda$ appearing in \eqref{eq:hstop_thomas}
are found in Table \ref{tab:fitting}. 
Again, a fit based on the original equation \eqref{eq:hstop} (or \eqref{eq:hstop_thomas}) rather
than Pouliquen's early fit \eqref{eq:hstop_exp} yields the best results.

For a flat bottom, such that $\lambda=0$, steady flow initiates and resides at or very tightly around
an inclination $\theta=12.5^\circ$ for all heights, see Fig.~\ref{fig:hstop_smooth}.
It is in agreement with the angle found in the laboratory experiments of \cite{GoujonThomasDalloz-Dubrujeaud2003}.
Hence, for a smooth base the flow is steady only at a single inclination, arrests for lower inclinations
and accelerates for larger inclinations. Such behaviour is in line with laboratory observations
and DPM simulations, {\em e.g.} \cite{VremanAlTaraziKuipersVanSintAnnalandBokhove2007}.
For $1/2<\lambda\leq2$, we observe Pouliquen-style behaviour in Fig.~\ref{fig:hstop_smooth}.
The angle $\delta_{1,\lambda}$ of flow initiation is nearly constant with respect to $\lambda$.
In contrast, the range of angles at which both steady and arrested flow is possible,
$\delta_{2,\lambda}-\delta_{1,\lambda}$, is maximal for $\lambda=1$ and decreases for smoother chutes with
$\lambda<1$, as follows from Table~\ref{tab:fitting}.
This has been reported in \cite{GoujonThomasDalloz-Dubrujeaud2003} for laboratory experiments,
who also observed a slight decrease of the interval $\delta_{2,\lambda}-\delta_{1,\lambda}$ for $\lambda>\lambda_c\approx 2$.
However, their $\lambda_c$ was measured for basal particles fixed at the same height and depended on the compactness of the base.
We observe a slight decrease of $\delta_{2,\lambda}$ for $\lambda=2$;
however, the fitting curves in Fig.~\ref{fig:hstop_smooth} do mildly overlap for $\lambda \ge 1$.

\begin{figure}[tbp]
%
\begin{psfrags}%
\psfragscanon%
%
\psfrag{s01}[t][t]{\fontsize{10}{15}\fontseries{m}\mathversion{normal}\fontshape{n}\selectfont \color[rgb]{0,0,0}\setlength{\tabcolsep}{0pt}\begin{tabular}{c}$\theta$\end{tabular}}%
\psfrag{s02}[b][b]{\fontsize{10}{15}\fontseries{m}\mathversion{normal}\fontshape{n}\selectfont \color[rgb]{0,0,0}\setlength{\tabcolsep}{0pt}\begin{tabular}{c}$h$\end{tabular}}%
\psfrag{s03}[r][r]{\fontsize{10}{15}\fontseries{m}\mathversion{normal}\fontshape{n}\selectfont \color[rgb]{0,0,0}\setlength{\tabcolsep}{0pt}\begin{tabular}{r}$\delta_{3,1/2}\rightarrow$\end{tabular}}%
\psfrag{s04}[r][r]{\fontsize{10}{15}\fontseries{m}\mathversion{normal}\fontshape{n}\selectfont \color[rgb]{0,0,0}\setlength{\tabcolsep}{0pt}\begin{tabular}{r}$\delta_{acc,1/2}\rightarrow$\end{tabular}}%
%
\fontsize{10}{15}\fontseries{m}\mathversion{normal}%
\fontshape{n}\selectfont%
%
\psfrag{x01}[t][t]{19}%
\psfrag{x02}[t][t]{20}%
\psfrag{x03}[t][t]{21}%
\psfrag{x04}[t][t]{22}%
\psfrag{x05}[t][t]{23}%
\psfrag{x06}[t][t]{24}%
\psfrag{x07}[t][t]{25}%
\psfrag{x08}[t][t]{26}%
\psfrag{x09}[t][t]{27}%
\psfrag{x10}[t][t]{28}%
\psfrag{x11}[t][t]{29}%
\psfrag{x12}[t][t]{30}%
%
\psfrag{v01}[r][r]{5}%
\psfrag{v02}[r][r]{10}%
\psfrag{v03}[r][r]{15}%
\psfrag{v04}[r][r]{20}%
\psfrag{v05}[r][r]{25}%
\psfrag{v06}[r][r]{30}%
\psfrag{v07}[r][r]{35}%
\psfrag{v08}[r][r]{40}%
\psfrag{v09}[r][r]{45}%
\psfrag{v10}[r][r]{50}%
%
\resizebox{8cm}{!}{\includegraphics{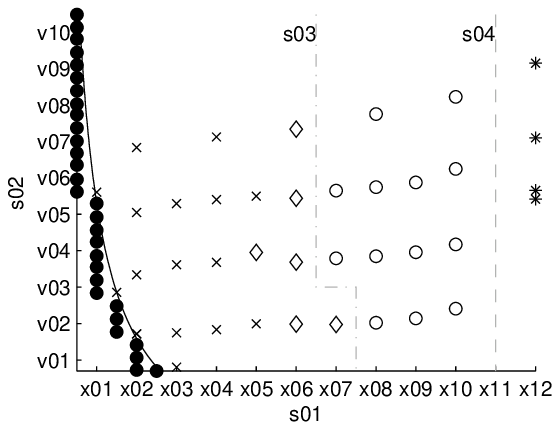}}%
\end{psfrags}%
%

	\smallskip
 	\scalebox{.5}{
%
\begin{psfrags}%
\psfragscanon%
%
\psfrag{s01}[t][t]{\fontsize{20}{30}\fontseries{m}\mathversion{normal}\fontshape{n}\selectfont \color[rgb]{0,0,0}\setlength{\tabcolsep}{0pt}\begin{tabular}{c}$z/h$\end{tabular}}%
\psfrag{s02}[b][b]{\fontsize{20}{30}\fontseries{m}\mathversion{normal}\fontshape{n}\selectfont \color[rgb]{0,0,0}\setlength{\tabcolsep}{0pt}\begin{tabular}{c}$\rho/\rho_p$\end{tabular}}%
\psfrag{s06}[][]{\fontsize{10}{15}\fontseries{m}\mathversion{normal}\fontshape{n}\selectfont \color[rgb]{0,0,0}\setlength{\tabcolsep}{0pt}\begin{tabular}{c} \end{tabular}}%
\psfrag{s07}[][]{\fontsize{10}{15}\fontseries{m}\mathversion{normal}\fontshape{n}\selectfont \color[rgb]{0,0,0}\setlength{\tabcolsep}{0pt}\begin{tabular}{c} \end{tabular}}%
%
\fontsize{20}{30}\fontseries{m}\mathversion{normal}%
\fontshape{n}\selectfont%
%
\psfrag{x01}[t][t]{0}%
\psfrag{x02}[t][t]{0.2}%
\psfrag{x03}[t][t]{0.4}%
\psfrag{x04}[t][t]{0.6}%
\psfrag{x05}[t][t]{0.8}%
\psfrag{x06}[t][t]{1}%
%
\psfrag{v01}[r][r]{0}%
\psfrag{v02}[r][r]{0.2}%
\psfrag{v03}[r][r]{0.4}%
\psfrag{v04}[r][r]{0.6}%
\psfrag{v05}[r][r]{0.8}%
%
\resizebox{8cm}{!}{\includegraphics{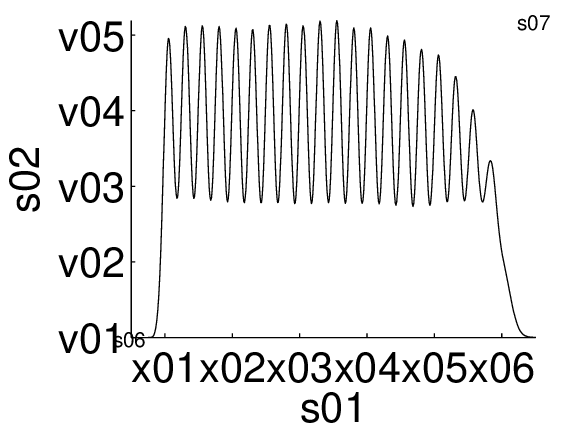}}%
\end{psfrags}%
%

%
\begin{psfrags}%
\psfragscanon%
%
\psfrag{s01}[t][t]{\fontsize{20}{30}\fontseries{m}\mathversion{normal}\fontshape{n}\selectfont \color[rgb]{0,0,0}\setlength{\tabcolsep}{0pt}\begin{tabular}{c}$t$\end{tabular}}%
\psfrag{s02}[b][b]{\fontsize{20}{30}\fontseries{m}\mathversion{normal}\fontshape{n}\selectfont \color[rgb]{0,0,0}\setlength{\tabcolsep}{0pt}\begin{tabular}{c}$E_{kin}/\langle E_{ela}\rangle$\end{tabular}}%
\psfrag{s06}[][]{\fontsize{10}{15}\fontseries{m}\mathversion{normal}\fontshape{n}\selectfont \color[rgb]{0,0,0}\setlength{\tabcolsep}{0pt}\begin{tabular}{c} \end{tabular}}%
\psfrag{s07}[][]{\fontsize{10}{15}\fontseries{m}\mathversion{normal}\fontshape{n}\selectfont \color[rgb]{0,0,0}\setlength{\tabcolsep}{0pt}\begin{tabular}{c} \end{tabular}}%
%
\fontsize{20}{30}\fontseries{m}\mathversion{normal}%
\fontshape{n}\selectfont%
%
\psfrag{x01}[t][t]{0}%
\psfrag{x02}[t][t]{500}%
\psfrag{x03}[t][t]{1000}%
\psfrag{x04}[t][t]{1500}%
\psfrag{x05}[t][t]{2000}%
%
\psfrag{v01}[r][r]{0.5}%
\psfrag{v02}[r][r]{1}%
\psfrag{v03}[r][r]{1.5}%
%
\resizebox{8cm}{!}{\includegraphics{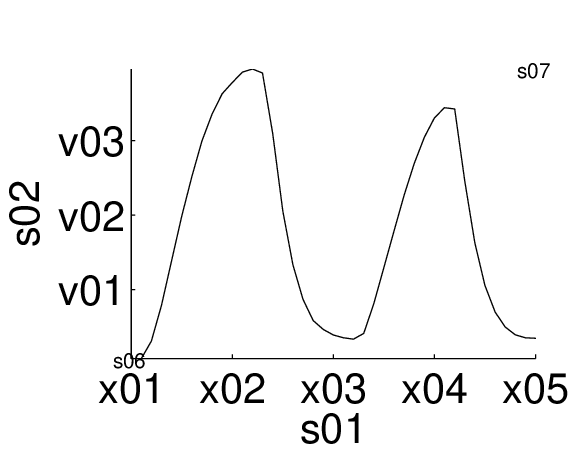}}%
\end{psfrags}%
%

	}
	\caption{Top: Overview of DPM simulations for $\lambda=1/2$, with markers denoting the flow state at $t=2000$:
	arrested $\bullet$, layered $\times$, oscillating $\diamond$, steady $\circ$, and accelerating $*$ flows.
	Demarcation line $h_{stop}(\theta;1/2)$ is fitted according to \eqref{eq:hstop}.
	Bottom left panel: Profile of particle volume fraction of layered flow at $H=20$, $\theta=22^\circ$. 
	Bottom right panel: Ratio of kinetic over mean elastic energy for oscillating flow at $H=30$,
	$\theta=24^\circ$. An example of oscillating flow is seen Fig.~\ref{fig:w_dependence_nu}.}
	\label{fig:hstop_diagram_2}
\end{figure}

\begin{table}[!b]
\centering
$$
\begin{array}{|l| |l|l|l| |l|l| |l|}
\hline
\lambda & \delta_{1,\lambda} & \delta_{2,\lambda} & A_\lambda & \beta_\lambda & \gamma_\lambda & \text{err}  \\\hline\hline
0 & 12.25 & 12.25 & - & 1.500 & -4.065 & 0.886  \\\hline
1/2 & 17.913 & 21.357 &11.959 & 0.300 & -0.236 & 0.229 \\\hline
2/3 & 17.217 & 24.302 &11.088 & 0.215 & -0.143 & 0.104  \\\hline
5/6 & 17.425 & 26.828 &7.465 & 0.203 & 0.039 & 0.113  \\\hline
1 & 17.708 & 32.780 &3.355 & 0.196 & 0.040 & 0.121  \\\hline
2 & 17.518 & 29.712 &5.290 & 0.189 & 0.080 & 0.137  \\\hline
\end{array}$$
\caption{Table of fitting parameters $\delta_{1,\lambda}$, $\delta_{2,\lambda}$, $A_{\lambda}$
for the curve $h_{stop}(\theta;\lambda)$ and $\beta_\lambda$,
$\gamma_\lambda$ for the flow rule \eqref{eq:froude_thomas}, including the variance of the flow rule,
$\text{err}(F-F_{data})$, for all steady flows.
}
\label{tab:fitting}
\end{table}

We recall that $\delta_{1,\lambda}$ and $\delta_{2,\lambda}$ are fitting parameters for the $h_{stop}$-curve \eqref{eq:hstop_thomas}
which does not necessarily imply, though it is expected, that the flow accelerates for angles greater than $\delta_{2,\lambda}$.
Surprisingly, while steady flow is observed exclusively for $\theta\in(\delta_{1,1},\delta_{2,1})$ when $\lambda=1$,
the range of angles associated with steady flow for smoother chutes ({\em i.e.}, when $\lambda<1$)
extends to greater inclinations with $\theta>\delta_{2,\lambda}$.
For these latter cases, $\delta_{acc,\lambda}>\delta_{2,\lambda}$ is defined as the smallest angle at which accelerating flow is observed;
the DPM simulations show that
\[
\delta_{acc,\lambda}=
\begin{cases}
	25^\circ\pm 1^\circ & \text{if }\lambda=0,\\
	29^\circ\pm1^\circ & \text{otherwise.}\label{deltacc}
\end{cases}
\]
Note that for the $\lambda=0$ case, between angles $12.5^\circ$ and $25^\circ$, the flow is steady and layered, because the friction factor is nonzero.

\def\lapprox{\ensuremath{\sim\kern-1em\raise 0.65ex\hbox{$&lt;$}}}
\def\rapprox{\ensuremath{\sim\kern-1em\raise 0.65ex\hbox{$&gt;$}}}

The above is illustrated in Fig.~\ref{fig:hstop_diagram_2} for $\lambda=1/2$,
where one observes the two different steady state regimes.
At higher angles, $\delta_{3,1/2}<\theta<\delta_{acc,1/2}$, a disordered regime similar to that for a rough base is observed. 
At smaller angles, $\delta_{1,1/2}<\theta<\delta_{3,1/2}$,
the flowing system self-organizes into a state of layered flow consisting of ordering in the $x$--$y$--plane for the bulk
(bottom left panel of Fig.~\ref{fig:hstop_diagram_2}),
except for a small intermediate region, $\theta\approx \delta_{3,1/2}$, where a transitional flow regime can be found.
It is characterized by large oscillations in the ratio of bulk averaged kinetic to elastic energy due to a spontaneous ordering and disordering,
or stop-and-go flow, of the system as a function of time (lower right panel).
The same flow regimes have been observed in \cite{SilbertGrestPlimptonLandry2002}, where the smoother
bottoms were achieved by arranging the base particles in a grid-like fashion.
In contrast, we always use of a fully disorded base and vary the roughness by changing the basal particle
size.

%
%
%
%
\section{Closure relations for the depth-averaged model}
\label{sec:closure}

The goal of this section is to close the shallow-layer equations \eqref{eq:swe}
by a determination of the basal friction $\mu$, the mean density $\bar{\rho}$, the stress ratio $K$,
and the velocity profile $\alpha$, using our DPMs.
A demarcation will be made of the flow regimes in which such a determination is possible.

\subsection{Friction $\mu$ of shallow-layer model}\label{sec:frictionmodel}

For the rough base several friction laws have been proposed, as detailed in Section \ref{sec:friction}.
In the following, we will compare these friction laws for the rough base,
$\lambda=1$, as well as for varying basal ratios $\lambda$.

\begin{figure*}[t]
\scalebox{.72}{
%
\begin{psfrags}%
\psfragscanon%
%
\psfrag{s01}[t][t]{\fontsize{14}{21}\fontseries{m}\mathversion{normal}\fontshape{n}\selectfont \color[rgb]{0,0,0}\setlength{\tabcolsep}{0pt}\begin{tabular}{c}$h/h_{stop}(\theta,\lambda=1)$\end{tabular}}%
\psfrag{s02}[b][b]{\fontsize{14}{21}\fontseries{m}\mathversion{normal}\fontshape{n}\selectfont \color[rgb]{0,0,0}\setlength{\tabcolsep}{0pt}\begin{tabular}{c}$F$\end{tabular}}%
%
\fontsize{14}{21}\fontseries{m}\mathversion{normal}%
\fontshape{n}\selectfont%
%
\psfrag{x01}[t][t]{0}%
\psfrag{x02}[t][t]{5}%
\psfrag{x03}[t][t]{10}%
\psfrag{x04}[t][t]{15}%
\psfrag{x05}[t][t]{20}%
%
\psfrag{v01}[r][r]{0}%
\psfrag{v02}[r][r]{0.5}%
\psfrag{v03}[r][r]{1}%
\psfrag{v04}[r][r]{1.5}%
\psfrag{v05}[r][r]{2}%
\psfrag{v06}[r][r]{2.5}%
\psfrag{v07}[r][r]{3}%
\psfrag{v08}[r][r]{3.5}%
\psfrag{v09}[r][r]{4}%
%
\resizebox{8cm}{!}{\includegraphics{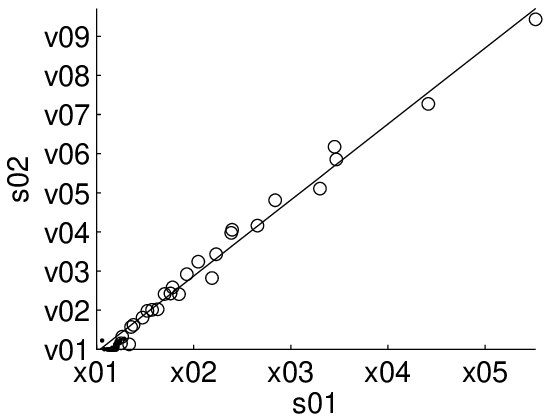}}%
\end{psfrags}%
%

%
\begin{psfrags}%
\psfragscanon%
%
\psfrag{s01}[t][t]{\fontsize{14}{21}\fontseries{m}\mathversion{normal}\fontshape{n}\selectfont \color[rgb]{0,0,0}\setlength{\tabcolsep}{0pt}\begin{tabular}{c}$h/h_{stop}(\theta,\lambda=1)$\end{tabular}}%
\psfrag{s02}[b][b]{\fontsize{14}{21}\fontseries{m}\mathversion{normal}\fontshape{n}\selectfont \color[rgb]{0,0,0}\setlength{\tabcolsep}{0pt}\begin{tabular}{c}$F$\end{tabular}}%
%
\fontsize{14}{21}\fontseries{m}\mathversion{normal}%
\fontshape{n}\selectfont%
%
\psfrag{x01}[t][t]{0}%
\psfrag{x02}[t][t]{5}%
\psfrag{x03}[t][t]{10}%
\psfrag{x04}[t][t]{15}%
\psfrag{x05}[t][t]{20}%
%
\psfrag{v01}[r][r]{0}%
\psfrag{v02}[r][r]{1}%
\psfrag{v03}[r][r]{2}%
\psfrag{v04}[r][r]{3}%
\psfrag{v05}[r][r]{4}%
\psfrag{v06}[r][r]{5}%
\psfrag{v07}[r][r]{6}%
\psfrag{v08}[r][r]{7}%
\psfrag{v09}[r][r]{8}%
\psfrag{v10}[r][r]{9}%
%
\resizebox{8cm}{!}{\includegraphics{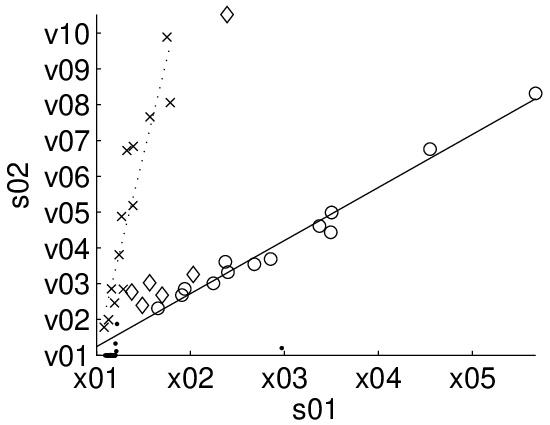}}%
\end{psfrags}%
%

%
\begin{psfrags}%
\psfragscanon%
%
\psfrag{s01}[t][t]{\fontsize{14}{21}\fontseries{m}\mathversion{normal}\fontshape{n}\selectfont \color[rgb]{0,0,0}\setlength{\tabcolsep}{0pt}\begin{tabular}{c}$h/h_{stop}(\theta,\lambda=1)$\end{tabular}}%
\psfrag{s02}[b][b]{\fontsize{14}{21}\fontseries{m}\mathversion{normal}\fontshape{n}\selectfont \color[rgb]{0,0,0}\setlength{\tabcolsep}{0pt}\begin{tabular}{c}$F$\end{tabular}}%
\psfrag{s05}[l][l]{\fontsize{12}{18}\fontseries{m}\mathversion{normal}\fontshape{n}\selectfont \color[rgb]{0,0,0}$\lambda=2$}%
\psfrag{s06}[l][l]{\fontsize{12}{18}\fontseries{m}\mathversion{normal}\fontshape{n}\selectfont \color[rgb]{0,0,0}$\lambda=0$}%
\psfrag{s07}[l][l]{\fontsize{12}{18}\fontseries{m}\mathversion{normal}\fontshape{n}\selectfont \color[rgb]{0,0,0}$\lambda=1/2$}%
\psfrag{s08}[l][l]{\fontsize{12}{18}\fontseries{m}\mathversion{normal}\fontshape{n}\selectfont \color[rgb]{0,0,0}$\lambda=2/3$}%
\psfrag{s09}[l][l]{\fontsize{12}{18}\fontseries{m}\mathversion{normal}\fontshape{n}\selectfont \color[rgb]{0,0,0}$\lambda=5/6$}%
\psfrag{s10}[l][l]{\fontsize{12}{18}\fontseries{m}\mathversion{normal}\fontshape{n}\selectfont \color[rgb]{0,0,0}$\lambda=1$}%
\psfrag{s11}[l][l]{\fontsize{12}{18}\fontseries{m}\mathversion{normal}\fontshape{n}\selectfont \color[rgb]{0,0,0}$\lambda=2$}%
\psfrag{s13}[][]{\fontsize{10}{15}\fontseries{m}\mathversion{normal}\fontshape{n}\selectfont \color[rgb]{0,0,0}\setlength{\tabcolsep}{0pt}\begin{tabular}{c} \end{tabular}}%
\psfrag{s14}[][]{\fontsize{10}{15}\fontseries{m}\mathversion{normal}\fontshape{n}\selectfont \color[rgb]{0,0,0}\setlength{\tabcolsep}{0pt}\begin{tabular}{c} \end{tabular}}%
%
\fontsize{12}{18}\fontseries{m}\mathversion{normal}%
\fontshape{n}\selectfont%
%
\psfrag{x01}[t][t]{0}%
\psfrag{x02}[t][t]{5}%
\psfrag{x03}[t][t]{10}%
\psfrag{x04}[t][t]{15}%
\psfrag{x05}[t][t]{20}%
%
\psfrag{v01}[r][r]{0}%
\psfrag{v02}[r][r]{2}%
\psfrag{v03}[r][r]{4}%
\psfrag{v04}[r][r]{6}%
\psfrag{v05}[r][r]{8}%
\psfrag{v06}[r][r]{10}%
%
\resizebox{8cm}{!}{\includegraphics{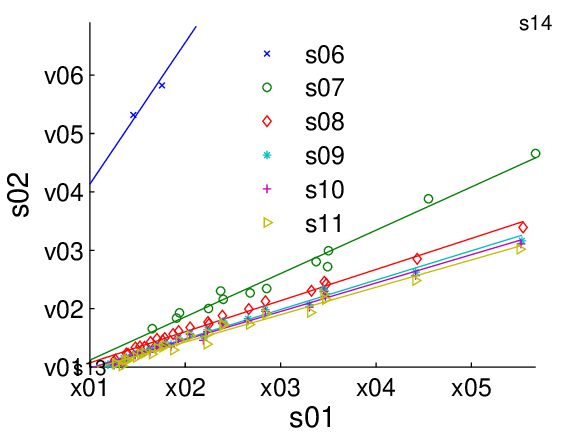}}%
\end{psfrags}%
%

}
\caption{
Froude number $F=u/\sqrt{gh}$ over height scaled by the stopping height for $\lambda=1$ (left),
$\lambda=1/2$ (centre), and for all basal roughnesses (right).
Data with symbols `$x$' denote steady layered, `$\circ$' steady, and `$\diamond$' oscillating flows.
Data with symbols `$.$' correspond to arrested or steady flows near $h_{stop}$.
The data is fit using $h_{stop}(\theta,\lambda=1)$ (solid lines). 
}
\label{fig:flowrule_smooth}
\end{figure*}

To obtain a function for the basal friction $\mu$, we used the approach of Pouliquen, who found that for a rough base
the Froude number is a linear function of $h/h_{stop}(\theta)$.
A first approach was to fit the Froude number to a linear function of $h/h_{stop}(\theta;\lambda)$
across the range of non-accelerating DPMs.
While this does work for $\lambda\ge 5/6$, a (linear or other) fit does not work for well for $\lambda\le 2/3$
because for the smoother bases layered and oscllating flows occur for
$\delta_{1,\lambda}<\theta_{stop}(h,\lambda)<\theta<\delta_{3,\lambda}$.
This is illustrated for $\lambda=1/2$ in Fig.~\ref{fig:hstop_diagram_2}.
Instead, the Froude number is fitted with $h_{stop}(\theta;1)$ such that
\begin{equation} \label{eq:froude_thomas}
F= \beta_\lambda \frac{h}{h_{stop}(\theta,1)} - \gamma_\lambda,\quad\mathrm{for}\quad \delta_{3,\lambda}<\theta<\delta_{acc,\lambda},
\end{equation}
where $\delta_{3,\lambda}$ is the largest angle at which oscillating flow is observed.
In Fig.~\ref{fig:hstop_diagram_2}, these angles $\delta_{acc,\lambda}$ and $\delta_{3,\lambda}$ are shown for the case $\lambda=1/2$.
Overall, the simulations reveal that
\[\delta_{3,\lambda}=\begin{cases}
	23^\circ\pm1^\circ & \text{if }\lambda=0,\\
	25.5^\circ\pm0.5^\circ & \text{if }\lambda=1/2\text{ and }H=10,\\
	24.5^\circ\pm0.5^\circ & \text{if }\lambda=1/2\text{ and }H>10,\\
	24.5^\circ\pm0.5^\circ & \text{if }\lambda=2/3\text{ and }H=10,\\
	\theta_{stop}(h;\lambda) & \text{otherwise.}\label{delta33}
\end{cases}\]
The results of such fits to the Pouliquen law for $\delta_{3,\lambda}<\theta<\delta_{acc,\lambda}$
are shown in Fig.~\ref{fig:flowrule_smooth}, with corresponding fitting parameters provided in Table \ref{tab:fitting}.
Shown is the Froude number $F= \bar{u}/\sqrt{g \cos\theta h}$ against the ratio of flow and stopping heights $h/h_{stop}(\theta;1)$.
For the disordered steady flow regime, concerning angles $\delta_{3,\lambda}<\theta<\delta_{acc,\lambda}$,
the data are seen to fit better with the stopping angle $h_{stop}(\theta;\lambda=1)$, the one for basal surface $\lambda=1$,
rather than with the actual stopping height $h_{stop}(\theta;\lambda)$. This is a key observation.
It shows that the Froude number $F$ increases as the roughness $\lambda$ decreases,
due to the lower dissipation at the base.
The weaker Froude number dependence for $\lambda=2$ seen in the right panel of Fig.~\ref{fig:flowrule_smooth}
is in line with the zero shear observed at the base in Fig.~\ref{fig:velocityX_lambda}.
The full set of fitting parameters and the standard error for the fit to \eqref{eq:froude_thomas}
are found in Table \ref{tab:fitting} with a standard error defined by
\begin{equation}
\text{err}(\{x_i\}_{i=1}^N)=\bigl(\sum_{i=1}^Nx_i^2/(N-1)\bigr)^{1/2}.
\end{equation} 
We remark that a fit to equation \eqref{eq:Pouliquen} is marginally better than Jenkins' adaption \eqref{eq:PJ},
but the differences are too small to discriminate accurately.

The situation for layered and oscillating flows is more complicated.
We illustrate that for the case $\lambda=1/2$.
Two fits are shown in  Fig.~\ref{fig:flowrule_smooth}, one for the layered case (dotted line concerning the crosses),
and one for the steady case (solid line concerning the circles).
The oscillating flows seem to defy a sensible fit because the flow swings irregularly between the layered and disordered states.
That oscillating behavior was also shown in Fig.~\ref{fig:hstop_diagram_2} (bottom right panel).

For steady flow, the shallow-layer equations \eqref{eq:swe} yield $\mu=\tan(\theta)$.
Indeed, within the range of steady or arrested flows DPMs 
confirm directly that the friction at the base lies within $|\mu-\tan^{-1}(\theta)|\linebreak[0] <0.4^\circ$.
In summary, for the steady flow regimes observed in our DPM simulations,
the friction coefficient of the depth-averaged equations \eqref{eq:swe} is parameterized to be
\begin{align}\nonumber 
\phantom{WW}\mu(h,F;\lambda) = \tan(\delta_{1,1}) + 
\frac{\tan(\delta_{2,1})-\tan(\delta_{1,1})}{\beta_\lambda h/(A_1 d(F+\gamma_\lambda))+1},\\
\quad\mathrm{for}\quad\delta_{3,\lambda}<\theta<\delta_{acc,\lambda}.
\label{eq:mu_thomas} 
\end{align}
where the parameters $\delta_{1,1},\ \delta_{2,1}, A_1$ are independent of the base; 
and, $\beta_\lambda$ and $\gamma_\lambda$ are depending explicitly on $\lambda$.
All values are found in Table \ref{tab:fitting}, with $\delta_{acc,\lambda}$ and $\delta_{3,\lambda}$ given in \eqref{deltacc} and \eqref{delta33}.
Despite its determination for steady flows, such a closure for $\mu$ is assumed and often observed to
be a reasonable `leading order' approach for unsteady shallow-layer flows.
Furthermore, for smoother bases, closure laws for layered and oscillating flows have eluded us.
It seems that the homogenization and steadiness assumptions of depth-averaged shallow-layer flow break down in these cases.

%
\subsection{Functions $\bar{\rho}, K, \alpha$ of shallow-layer model}\label{sec:depthaveraging}
%

DPM simulations of steady uniform flows are considered
for disordered steady flow with $\delta_3^\lambda<\theta\leq\delta_{acc}^\lambda$,
to determine closures for $\bar\rho$, $K$ and $\alpha$ as functions of continuum fields $\bar{u}$ and $h$.
The layered and oscillating flow regimes are thus excluded momentarily.

\begin{figure}[tbp]
%
\begin{psfrags}%
\psfragscanon%
%
\psfrag{s01}[t][t]{\fontsize{10}{15}\fontseries{m}\mathversion{normal}\fontshape{n}\selectfont \color[rgb]{0,0,0}\setlength{\tabcolsep}{0pt}\begin{tabular}{c}$\theta$\end{tabular}}%
\psfrag{s02}[b][b]{\fontsize{10}{15}\fontseries{m}\mathversion{normal}\fontshape{n}\selectfont \color[rgb]{0,0,0}\setlength{\tabcolsep}{0pt}\begin{tabular}{c}$\bar\rho/\rho_p$\end{tabular}}%
\psfrag{s05}[l][l]{\fontsize{10}{15}\fontseries{m}\mathversion{normal}\fontshape{n}\selectfont \color[rgb]{0,0,0}$H=10$}%
\psfrag{s06}[l][l]{\fontsize{10}{15}\fontseries{m}\mathversion{normal}\fontshape{n}\selectfont \color[rgb]{0,0,0}$\bar{\rho}_c/\rho_p$}%
\psfrag{s07}[l][l]{\fontsize{10}{15}\fontseries{m}\mathversion{normal}\fontshape{n}\selectfont \color[rgb]{0,0,0}$\bar{\rho}_c^{fit}/\rho_p$}%
\psfrag{s08}[l][l]{\fontsize{10}{15}\fontseries{m}\mathversion{normal}\fontshape{n}\selectfont \color[rgb]{0,0,0}$H=40$}%
\psfrag{s09}[l][l]{\fontsize{10}{15}\fontseries{m}\mathversion{normal}\fontshape{n}\selectfont \color[rgb]{0,0,0}$H=30$}%
\psfrag{s10}[l][l]{\fontsize{10}{15}\fontseries{m}\mathversion{normal}\fontshape{n}\selectfont \color[rgb]{0,0,0}$H=20$}%
\psfrag{s11}[l][l]{\fontsize{10}{15}\fontseries{m}\mathversion{normal}\fontshape{n}\selectfont \color[rgb]{0,0,0}$H=10$}%
\psfrag{s13}[][]{\fontsize{10}{15}\fontseries{m}\mathversion{normal}\fontshape{n}\selectfont \color[rgb]{0,0,0}\setlength{\tabcolsep}{0pt}\begin{tabular}{c} \end{tabular}}%
\psfrag{s14}[][]{\fontsize{10}{15}\fontseries{m}\mathversion{normal}\fontshape{n}\selectfont \color[rgb]{0,0,0}\setlength{\tabcolsep}{0pt}\begin{tabular}{c} \end{tabular}}%
%
\fontsize{10}{15}\fontseries{m}\mathversion{normal}%
\fontshape{n}\selectfont%
%
\psfrag{x01}[t][t]{22}%
\psfrag{x02}[t][t]{23}%
\psfrag{x03}[t][t]{24}%
\psfrag{x04}[t][t]{25}%
\psfrag{x05}[t][t]{26}%
\psfrag{x06}[t][t]{27}%
\psfrag{x07}[t][t]{28}%
%
\psfrag{v01}[r][r]{0.51}%
\psfrag{v02}[r][r]{0.52}%
\psfrag{v03}[r][r]{0.53}%
\psfrag{v04}[r][r]{0.54}%
\psfrag{v05}[r][r]{0.55}%
\psfrag{v06}[r][r]{0.56}%
\psfrag{v07}[r][r]{0.57}%
%
\resizebox{8cm}{!}{\includegraphics{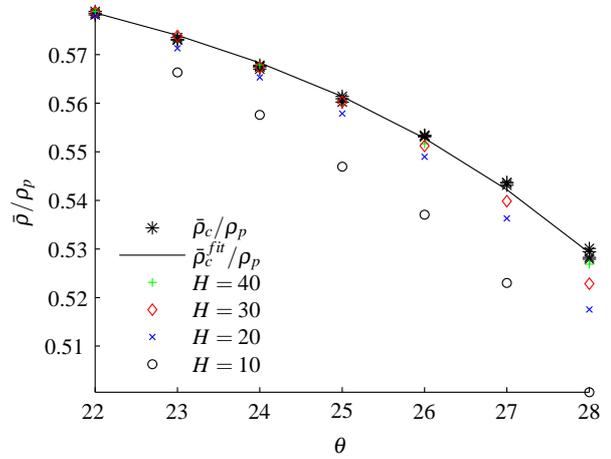}}%
\end{psfrags}%
%

\caption{Mean volume fraction $\bar\rho/\rho_p$ for roughness $\lambda=1$, and varying approximated heights $H$
and inclinations $\theta$. The mean volume fraction in the bulk, $\bar{\rho}/\rho_p$, denoted by $*$, collapses onto
a function of the inclination (solid line), while it shows a small
dependence on the flow height, due to the density decrease near base and surface.}
\label{fig:Nu}
\end{figure}

All steady disordered flows show a constant density profile in the bulk of the flow,
{\em cf}. Fig.~\ref{fig:w_dependence_nu}, while the density decreases near the base and the surface.
The lower density region at the base spans about two particle diameters for $\lambda>0$, and less
than $9d$ for $\lambda=0$, while the surface region spans always less than $4d$.
Thus, a mean bulk density can roughly be defined as
\begin{equation}
\bar{\rho}_c = \frac{1}{h-6d} \int_{b+2d}^{s-4d} \rho(z) \,dz.
\end{equation}
In Fig.~\ref{fig:Nu}, the bulk volume fraction and the mean volume fraction are shown for roughness $\lambda=1$ and
varying height and inclination. The bulk volume fraction decreases with inclination $\theta$, but is independent of
flow height and roughness, whereas the mean volume fraction depends also on flow height and roughness.
We fit the mean bulk density of all steady disordered flows with $\lambda>0$ to an arbitrary function
\begin{subequations}
\begin{equation}
\bar{\rho}^{fit}_c/\rho_p = c_0-\exp{(\theta-c_2)/c_1},
\end{equation} 
with fitting parameters
\begin{equation}
c_0=0.5985, c_1=4.8^{\circ},\text{and}\, c_2=40.85^{\circ}.
\end{equation} 
Standard deviations of the mean bulk volume fraction and mean volume fraction
for all cases with $\lambda>0$ are
\begin{equation}
\text{err}(\bar{\rho}_c^{fit}-\bar{\rho}_c)=0.002,\quad\textrm{and}\quad \text{err}(\bar{\rho}_c^{fit}-{\rho})=0.018. 
\end{equation} 
\end{subequations}

Secondly, the normal stress ratio $K=\bar\sigma_{xx}/\bar\sigma_{zz}$ is determined. It describes the anisotropy
of the stress tensor and is expected to be unity under isotropic conditions.
The range of $K$ for steady disordered flow is generally small, ranging from 0.98 to 1.07,
except for $\lambda=0$, where it can be as low as 0.68.
The stress anisotropy generally increases with inclination.
For $\lambda>0$, $K$ fits to a function linear in $\theta$,
\begin{equation}
K^{fit}=1+(\theta-d_1)/d_0
\end{equation}
with $d_0=132^{\circ}$ and $d_1=21.50^\circ$.
The model results give a small standard error of $\text{err}(K-K^{fit})=0.013$.
Given that the dependence on inclination is small, we may as well take $K\approx 1$.

\begin{figure*}[thbp]
\centering
%
\begin{psfrags}%
\psfragscanon%
%
\psfrag{s01}[t][t]{\fontsize{10}{15}\fontseries{m}\mathversion{normal}\fontshape{n}\selectfont \color[rgb]{0,0,0}\setlength{\tabcolsep}{0pt}\begin{tabular}{c}$h_{stop}(\theta;1)/h$\end{tabular}}%
\psfrag{s02}[b][b]{\fontsize{10}{15}\fontseries{m}\mathversion{normal}\fontshape{n}\selectfont \color[rgb]{0,0,0}\setlength{\tabcolsep}{0pt}\begin{tabular}{c}$b_{1}$\end{tabular}}%
\psfrag{s05}[l][l]{\fontsize{10}{15}\fontseries{m}\mathversion{normal}\fontshape{n}\selectfont \color[rgb]{0,0,0}$H=10$}%
\psfrag{s06}[l][l]{\fontsize{10}{15}\fontseries{m}\mathversion{normal}\fontshape{n}\selectfont \color[rgb]{0,0,0}$fit$}%
\psfrag{s07}[l][l]{\fontsize{10}{15}\fontseries{m}\mathversion{normal}\fontshape{n}\selectfont \color[rgb]{0,0,0}$H=40$}%
\psfrag{s08}[l][l]{\fontsize{10}{15}\fontseries{m}\mathversion{normal}\fontshape{n}\selectfont \color[rgb]{0,0,0}$H=30$}%
\psfrag{s09}[l][l]{\fontsize{10}{15}\fontseries{m}\mathversion{normal}\fontshape{n}\selectfont \color[rgb]{0,0,0}$H=20$}%
\psfrag{s10}[l][l]{\fontsize{10}{15}\fontseries{m}\mathversion{normal}\fontshape{n}\selectfont \color[rgb]{0,0,0}$H=10$}%
\psfrag{s12}[][]{\fontsize{10}{15}\fontseries{m}\mathversion{normal}\fontshape{n}\selectfont \color[rgb]{0,0,0}\setlength{\tabcolsep}{0pt}\begin{tabular}{c} \end{tabular}}%
\psfrag{s13}[][]{\fontsize{10}{15}\fontseries{m}\mathversion{normal}\fontshape{n}\selectfont \color[rgb]{0,0,0}\setlength{\tabcolsep}{0pt}\begin{tabular}{c} \end{tabular}}%
%
\fontsize{10}{15}\fontseries{m}\mathversion{normal}%
\fontshape{n}\selectfont%
%
\psfrag{x01}[t][t]{0}%
\psfrag{x02}[t][t]{0.1}%
\psfrag{x03}[t][t]{0.2}%
\psfrag{x04}[t][t]{0.3}%
\psfrag{x05}[t][t]{0.4}%
\psfrag{x06}[t][t]{0.5}%
\psfrag{x07}[t][t]{0.6}%
\psfrag{x08}[t][t]{0.7}%
%
\psfrag{v01}[r][r]{0}%
\psfrag{v02}[r][r]{0.5}%
\psfrag{v03}[r][r]{1}%
\psfrag{v04}[r][r]{1.5}%
\psfrag{v05}[r][r]{2}%
\psfrag{v06}[r][r]{2.5}%
\psfrag{v07}[r][r]{3}%
\psfrag{v08}[r][r]{3.5}%
\psfrag{v09}[r][r]{4}%
\psfrag{v10}[r][r]{4.5}%
%
\resizebox{8cm}{!}{\includegraphics{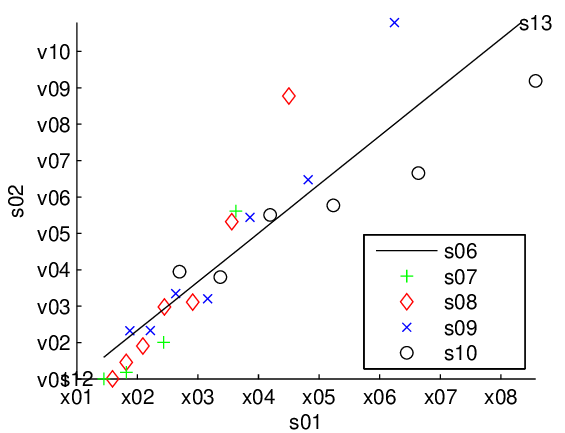}}%
\end{psfrags}%
%

%
\begin{psfrags}%
\psfragscanon%
%
\psfrag{s01}[t][t]{\fontsize{10}{15}\fontseries{m}\mathversion{normal}\fontshape{n}\selectfont \color[rgb]{0,0,0}\setlength{\tabcolsep}{0pt}\begin{tabular}{c}$\theta$\end{tabular}}%
\psfrag{s02}[b][b]{\fontsize{10}{15}\fontseries{m}\mathversion{normal}\fontshape{n}\selectfont \color[rgb]{0,0,0}\setlength{\tabcolsep}{0pt}\begin{tabular}{c}$\alpha$\end{tabular}}%
%
\fontsize{10}{15}\fontseries{m}\mathversion{normal}%
\fontshape{n}\selectfont%
%
\psfrag{x01}[t][t]{22}%
\psfrag{x02}[t][t]{23}%
\psfrag{x03}[t][t]{24}%
\psfrag{x04}[t][t]{25}%
\psfrag{x05}[t][t]{26}%
\psfrag{x06}[t][t]{27}%
\psfrag{x07}[t][t]{28}%
%
\psfrag{v01}[r][r]{1.25}%
\psfrag{v02}[r][r]{1.3}%
\psfrag{v03}[r][r]{1.35}%
\psfrag{v04}[r][r]{1.4}%
\psfrag{v05}[r][r]{1.45}%
\psfrag{v06}[r][r]{1.5}%
%
\resizebox{8cm}{!}{\includegraphics{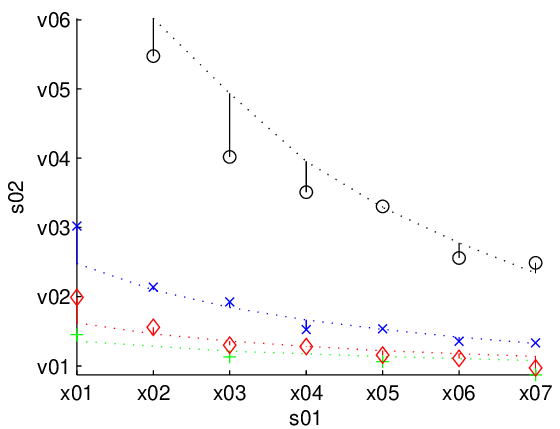}}%
\end{psfrags}%
%

\caption{
Left figure: fitting parameter $b_\lambda$ as a function of $h_{stop}(\theta;1)/h$ for $\lambda=1$ and varying height
$h$ and inclination $\theta$. The solid line shows the linear fit used to obtain $\alpha$ from equations \eqref{eq:alpha}.
Right: Shape factor $\alpha$ for $\lambda=1$ and varying height $h$ and inclination $\theta$.
Markers denote the simulation data, while dotted lines denote fits using \eqref{eq:alpha} with corresponding
coefficients from Table \ref{tab:alpha}. Fitted values and simulation data are connected by a solid line.
}
\label{fig:alpha}
\end{figure*} 

\begin{figure}[thbp]
\centering
%
\begin{psfrags}%
\psfragscanon%
%
\psfrag{s01}[t][t]{\fontsize{10}{15}\fontseries{m}\mathversion{normal}\fontshape{n}\selectfont \color[rgb]{0,0,0}\setlength{\tabcolsep}{0pt}\begin{tabular}{c}$(z-b)/h$\end{tabular}}%
\psfrag{s02}[b][b]{\fontsize{10}{15}\fontseries{m}\mathversion{normal}\fontshape{n}\selectfont \color[rgb]{0,0,0}\setlength{\tabcolsep}{0pt}\begin{tabular}{c}$h/\bar{u}\partial_zu$\end{tabular}}%
\psfrag{s05}[l][l]{\fontsize{10}{15}\fontseries{m}\mathversion{normal}\fontshape{n}\selectfont \color[rgb]{0,0,0}fit $\lambda=2/3$}%
\psfrag{s06}[l][l]{\fontsize{10}{15}\fontseries{m}\mathversion{normal}\fontshape{n}\selectfont \color[rgb]{0,0,0}$\lambda=1$}%
\psfrag{s07}[l][l]{\fontsize{10}{15}\fontseries{m}\mathversion{normal}\fontshape{n}\selectfont \color[rgb]{0,0,0}$\lambda=2/3$}%
\psfrag{s08}[l][l]{\fontsize{10}{15}\fontseries{m}\mathversion{normal}\fontshape{n}\selectfont \color[rgb]{0,0,0}$\lambda=0$}%
\psfrag{s09}[l][l]{\fontsize{10}{15}\fontseries{m}\mathversion{normal}\fontshape{n}\selectfont \color[rgb]{0,0,0}fit $\lambda=1$}%
\psfrag{s10}[l][l]{\fontsize{10}{15}\fontseries{m}\mathversion{normal}\fontshape{n}\selectfont \color[rgb]{0,0,0}fit $\lambda=2/3$}%
\psfrag{s12}[][]{\fontsize{10}{15}\fontseries{m}\mathversion{normal}\fontshape{n}\selectfont \color[rgb]{0,0,0}\setlength{\tabcolsep}{0pt}\begin{tabular}{c} \end{tabular}}%
\psfrag{s13}[][]{\fontsize{10}{15}\fontseries{m}\mathversion{normal}\fontshape{n}\selectfont \color[rgb]{0,0,0}\setlength{\tabcolsep}{0pt}\begin{tabular}{c} \end{tabular}}%
%
\fontsize{10}{15}\fontseries{m}\mathversion{normal}%
\fontshape{n}\selectfont%
%
\psfrag{x01}[t][t]{0}%
\psfrag{x02}[t][t]{0.1}%
\psfrag{x03}[t][t]{0.2}%
\psfrag{x04}[t][t]{0.3}%
\psfrag{x05}[t][t]{0.4}%
\psfrag{x06}[t][t]{0.5}%
\psfrag{x07}[t][t]{0.6}%
\psfrag{x08}[t][t]{0.7}%
\psfrag{x09}[t][t]{0.8}%
\psfrag{x10}[t][t]{0.9}%
\psfrag{x11}[t][t]{1}%
%
\psfrag{v01}[r][r]{0}%
\psfrag{v02}[r][r]{0.5}%
\psfrag{v03}[r][r]{1}%
\psfrag{v04}[r][r]{1.5}%
\psfrag{v05}[r][r]{2}%
\psfrag{v06}[r][r]{2.5}%
\psfrag{v07}[r][r]{3}%
\psfrag{v08}[r][r]{3.5}%
%
\resizebox{8cm}{!}{\includegraphics{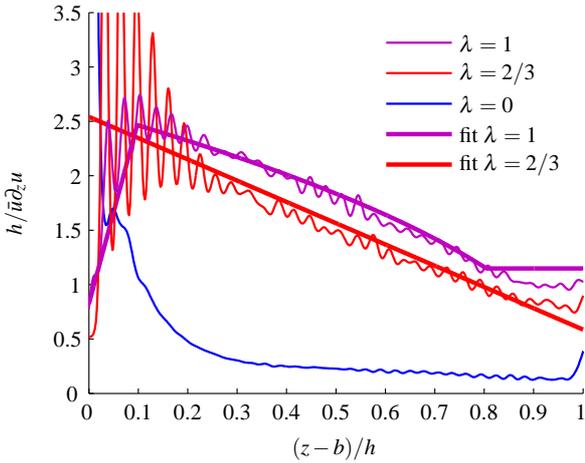}}%
\end{psfrags}%
%
 
\caption{
Depth profile of normalized strain, $(h/\bar{u})\,\partial_z u$ corresponding to velocity profiles shown in
Fig.~\ref{fig:velocityX_lambda}. For rough bases, the strain is modelled by a Bagnold profile, except near the base and surface.
For smoother bases, $\lambda\leq2/3$, the profile becomes linear. For $\lambda=0$, a large slip velocity is observed,
and the strain becomes inversely proportional to the depth.}
\label{fig:strain}
\end{figure} 

Finally, we develop a fit for the shape factor $\alpha(\lambda)=\bar{u}^2/\overline{u^2}$.
The fit is based on a phenomenological model of the observed velocity profiles, as shown in Fig.~\ref{fig:velocityX_lambda}. 
For rough bases $\lambda\geq5/6$, a Bagnold velocity profile,
\begin{subequations}\label{eq:alpha}
\begin{equation}
u_B(z)=\frac{5}{3}\bar{u} \left(1-\bigl({(h-z)}/{h}\bigr)^{3/2}\right),
\end{equation}
is observed in the bulk of the flow; a linear profile in the surface layer, which is about $5d$ thick;
and, a convex profile with no slip in the base layer, whose thickness $b_\lambda$ increases as the height approaches the stopping height.
No kinks occur at the intersection of the layers. 
Thus, we model the velocity by
\begin{gather}
\pfrac{u}{z}(z;b_\lambda)=
\left\{
\begin{array}{cc}
\pfrac{u_B}{z}(z=b_\lambda)(1-\frac{2}{3}\frac{b_\lambda-z}{b_\lambda-b}), & z<b_\lambda,\\
\pfrac{u_B}{z}(z), & \hspace{-1.4cm}b_\lambda\leq z<\max(s-5,b_\lambda),\hspace*{-1in}\\
\pfrac{u_B}{z}\max(s-5,b_\lambda), & \text{otherwise}, 
\end{array}
\right. \nonumber\\
u(0;b_\lambda)=0\text{ for } \lambda\geq5/6.
\end{gather}
A fit to the strain $\partial_z u$ in Fig.~\ref{fig:strain} shows it is well approximated by this model.
The parameter $b_\lambda$ decreases with increasing distance from the stopping height, and a simple fit reads
\begin{equation}
b_\lambda=d_\lambda {h_{stop}(\theta,1)}/{h},
\end{equation}
\end{subequations}
where $h_{stop}(\theta,1)$ was chosen since $h_{stop}(\theta,\lambda)$ does not provide values for all inclinations for which steady flow is observed. 
Subsequently, the shape factor $\alpha(\lambda)=\bar{u}^2/\overline{u^2}$ can be computed numerically and compared to
the measured values in Fig.~\ref{fig:alpha}. 
The coefficients $b_\lambda$ are given in Table \ref{tab:alpha}. 

For $\lambda\leq2/3$, the dependence of the shape factor on height and inclination diminishes and can be
approximated with a constant value $\alpha(\lambda)$. The Bagnold profile disappears and the flow becomes more convex and plug-like,
as shown in Fig.~\ref{fig:velocityX_lambda}. Each velocity profile will be analyzed in turn next.

For $\lambda=0$, the slip is so large that we can assume plug flow to hold.
There is almost no slip for $\lambda=2/3$ and a slip of approximately $u(0)/\max_z u(z)=1/6$ for $\lambda=1/2$.
We neglect the variations at the surface and the bulk and model.
Thus, we model the velocity profiles as
\begin{gather}
\frac{u(z)}{\bar{u}}=
\begin{cases}
5/3-(1-z/h)^2, &\lambda=2/3,\\
0.16+0.84(5/3-(1-z/h)^2), &\lambda=1/2,\\
1 , &\lambda=0,
\end{cases}
\end{gather}
The corresponding coefficients $\alpha(\lambda)$ are found in Table \ref{tab:alpha} and provide a good fit to the data. 

\begin{table}[bt]
$$\begin{array}{|l| |l|l| |l|l|r|r|r|r|r|r|r|}\hline
\lambda & \alpha(\lambda) & b_\lambda & \text{err}\\\hline\hline
0 & 1.000 & & 0.012\\ 
1/2 & 1.142 & & 0.01919\\
2/3 & 1.201 & & 0.02741\\
5/6 & & 5.105 & 0.02558\\
1 & & 6.678 & 0.01763\\
2 & & 13.75 & 0.04479\\
\hline\end{array}$$
\caption{Fitting for the shape factor $\alpha=\alpha(\lambda)$ for $\lambda\leq2/3$ and $\alpha=\bar{u}^2/\overline{u^2}$,
$u=u(z;b_\lambda)$ for $\lambda\geq 5/6$, and the standard error.
Closure relations are fitted to all data sets of steady unordered flow, $\delta_3^\lambda<\theta\leq\delta_{acc}^\lambda$.}
\label{tab:alpha}
\end{table}

In summary, the functions $\bar\rho$, $\alpha$ and $K$ depend on the inclination $\theta$ and the height $h$.
The inclination $\theta$ in turn can be written as a function of the friction coefficient $\mu$ such that
$\theta=\tan^{-1}(\mu(h,\bar{u}))$. This allows us to describe the parameters of the shallow-layer model
in terms of the height $h$, roughness $\lambda$, and friction $\mu(h,\bar{u})$
and thus provides a closure proper for the system.
The different behavior for the varying $\lambda$'s remains an open issue,
since we only provided emperical fits above.


\section{Conclusion}
\label{sec:conclusion}

\subsection{Summary}

In this article, an extensive series of DPM simulations was used to determine
closure relations for the popular shallow-layer model of granular flows on inclined chutes.
The latter model is a depth-averaged continuum model with a macro-scale
variable thickness $h=h(x,t)$ and with a mean velocity $\bar{u}=\bar{u}(x,t)$ as variables.
For simplicity, we assumed uniformity in the lateral $y$--direction.
The flow consisted of monodispersed particles of diameter $d$ and the base of monodispersed particles of
diameter $\lambda d$. The bottom roughness, or diameter ratio, $\lambda$ was systematically varied.
Simulations revealed the existence of a range of chute inclinations $\theta$
for which horizontal and temporal variations are small enough to produce an approximately steady and uniform flow.
Particle flows with variations in height $h$ and inclination $\theta$ were numerically investigated
for varying basal roughness $\lambda$.

We observed the following phenomenology:
at small inclinations, the flow quickly forms a static pile, while at large inclinations,
the flow continues to accelerate. Between these two regimes there was a range of
inclinations in which steady flows occurred ({\em cf.}, Fig.~\ref{fig:silbert_comparison}).
The curve $h_{stop}(\theta;\lambda)$, a function of height versus inclination,
forms the separatrix between arrested and steady flow, as
a function of basal roughness (Figs.~\ref{fig:silbert_comparison} and \ref{fig:hstop_smooth}).
For smaller basal roughness, steady states arise at smaller inclinations and heights, and the range of angles shrinks
for which steady flow is possible. Other types of steady flow were observed at small inclinations
for small base particles, showing a strong layering in depth as well as disordered and oscillatory flows
({\em cf.}, Fig.~\ref{fig:hstop_diagram_2}).

Depth profiles for density, velocity and stress were constructed using coarse-grained macroscopic fields.
The coarse-graining width was carefully chosen to preserve some microscopic structure as well as macroscopic gradients
({\em cf.}, Fig.~\ref{fig:w_dependence_nu}).
The assumptions of depth-averaged theory were confirmed in the simulations
for a certain range of steady, uniform flows:
the density was constant at depth, and the downward normal stress as well as shear stress were lithostatic.
A often-used key assumption, often used, is that statistically steady DPM simulations, or laboratory experiments,
are relevant to find the closure relations even for time-dependent continuum shallow-layer models.

Consequently, four closure relations could in principle be determined:
for basal friction $\mu$, stress ratio $K$, mean density $\bar{\rho}$,
and for the shape of the velocity profile $\alpha$.
Firstly, basal friction $\mu=\tan\theta$ was shown to be a function of height and
flow velocity (Table \ref{tab:fitting}). 
Pouliquen's approach was found to be valid, with the Froude number as a linear function of $h/h_{stop}(\theta;\lambda=1)$.
This fitting approach was extended to smoother cases with $\lambda<1$,
where the Froude number was fitted to $h/h_{stop}(\theta;\lambda=1)$
instead of the $h/h_{stop}(\theta;\lambda)$ for the actual $\lambda$ or basal roughness.
The stopping curve associated with the diameter $\lambda=1$ of the flowing particles
is more relevant than the stopping with the actual $\lambda$.
One possible explanation is that there is a boundary layer of intermittently
slow flow particles that originated in the bulk, and that shields the smoother base from the bulk flow.
Closure relations for the mean density $\bar\rho$, stress anisotropy $K$ and shape factor $\alpha$ were also established
as follows.
For rough bases with $\lambda\ge 5/6$, the determined closures were valid
for $\theta_{stop}(h;\lambda)=\delta_{3,\lambda}<\theta<\delta_{acc,\lambda}$,
with $\theta_{stop}(h;\lambda)$ the inverse of the $h_{stop}(\theta;\lambda)$--curve between arrested and dynamic flow.
For smaller roughnesses with $\lambda\le 2/3$ and $\theta_{stop}(h;\lambda)<\theta<\delta_{3,\lambda}$, layered and oscillating flows arose for which we are (as yet) unable to capture closures.
For these smoother bases with $\lambda\le 5/6$, closures were obtained
for the range $\delta_{3,\lambda}<\delta<\delta_{acc,\lambda}$.

\subsection{Open questions}

What does the granular shallow-layer model enable us to do, and what can we not do with it?
In the range of steady flows, this continuum model can be used to predict steady and time-dependent flows.
Strictly speaking, this is only allowed for steady flow in the established inclination range
$\delta_{3,\lambda}<\theta<\delta_{acc,\lambda}$, but
it can be expected to remain valid for the slowly-varying dynamic cases as well.
It is often the case, however, that even rapidly-changing flows can be captured
by models that should only be valid for the slowly-varying cases.
Consequently, a systematic study of the validity of the shallow-layer model is required.
By respectively extending
the ``hydraulic'' analysis for fluidized granular matter and water in Vreman {\em et al.}
\cite{VremanAlTaraziKuipersVanSintAnnalandBokhove2007} and Akers and Bokhove \cite{akersb2008},
granular flows within constrictions become a nice, analytically-treatable targets.
Such flows in constrictions reach a steady state and appear (partially) accessible by direct DPM simulations.

What do these results enable us to do?
Whether the steady DPM-based closures are valid
across granular ``hydraulic'' jumps in such steady and constricting flows is of interest.
Whether the steady DPM-based closures hold for (slow) transient routes towards such steady states is of interest, too.
What closures should be used outside the formal range of applicability for the smoother bases,
so for the layered and oscillating flows for
$\theta_{stop}(h;\lambda)<\theta<\delta_{3,\lambda}$ and the accelerating flows for $\theta>\delta_{acc,\lambda}$,
appears a tantalising, and as yet, open question.

What are we not able do?
Although, we did observe layered and oscillating flows in our DPM simulations, it is doubtful
as to whether the homogenization assumption that led to the shallow-layer model is sufficient.
Nonetheless, the lithostatic balance relation is shown to hold for the DPM simulations,
as expected from standard asymptotic analysis using the aspect ratio of normal to planar velocity and length scales.

\subsection{Outlook}

Alternatively, a multi-scale modelling approach might be adopted such as the
heterogeneuous, multiscale methodology \cite{WeinanEngquistBjornLiRenVandern-Eijnden2007}, among others,
in which closure relations for discretisations ({\em e.g.}, \cite{PeschBellSollieAmbatiBokhoveVegt2007})
depth-averaged shallow-layer models are coupled to DPM simulations in selected regions in space and time.
Thus computational costs would be diminished while accurate closure relations
are gathered intermittently in time and space.

For future work, we advocate the extension of our DPM simulations with investigation of the three-dimensional closure relations.
We surmise that reduced lithostatic models for shallow granular flows could be more consistently derived
from three-dimensional continuum models with stress closure determined from DPM simulations
in combination with laboratory measurements.
These new models would be reduced and therefore computationally still manageable for large-scale
debris flows; for example, when the degrees of freedom in the vertical remain limited, but are extended beyond only one degree of freedom.
Such reduced modelling is akin to hydrostatic modelling in water-wave and coastal hydrodynamics.

\bibliographystyle{alpha}
\bibliography{paper,book}

\section*{Acknowledgements}
The authors would like to thank the Institute of Mechanics, Processes and Control, Twente (IMPACT) for its financial support.

\newpage
\appendix
%
\section{System of Hamiltonian equations in the dissipation-free limit}
\label{sec:hamiltonian}

The purpose of this appendix is to show that the particle system in the dissipation- and yield-free limit,
\emph{i.e.}, $\gamma^{\,n}=\gamma^{\,t}=0$ and $\mu_c\to\infty$, is a Hamiltonian system. 
It subsequently facilitates the derivation of so-called conservative or symplectic
discretization schemes, in time. These have been shown and are believed to provide better
long-term statistics than classical time discretization schemes.
Furthermore, analysis of the Hamiltonian limit allows one to clearly demarcate the transfer
of energy between its kinetic, elastic, and internal components.

We show that the normal and tangential forces are elastic, that is the system does not dissipate energy;
instead kinetic energy is converted into potential energy in the springs and vice versa.
If the tangential spring is not fully unloaded when two particles loose contact,
the potential energy stored in the tangential spring is converted into internal potential energy in each particle (vibrations).

We use the notation given in \S \ref{sec:contactlaw}. In the dissipation- and yield-free limit,
the contact force between particles $i,j$ is given by
\[ \vec{f}_{ij} = k^n \delta_{ij}^n \hat{\vec{n}}_{ij} - k^{\,t} \pmb{\delta}_{ij}^t, \label{eq:forceHam}\]
where $\delta_{ij}^n$ and $\pmb{\delta}_{ij}^t$ are given by equations \eqref{eq:deltan} and \eqref{eq:deltat}.
The equations of motion for translational and angular momentum of particle $i$ are given by, 
\begin{subequations}
\begin{align}
\frac{d}{dt}\vec{r}_i =& \vec{p}_i/m_i,\quad \frac{d}{dt}\svec{\alpha}_i = \svec{\phi}_i\\
\\
\frac{d}{dt}\vec{p}_i =& m_i \vec{g} + \sum_{j\not=i} \vec{f}_{ij},\quad 
\frac{d}{dt}\svec{\phi}_i  = \sum_{j\not=i} \vec{b}_{ij} \times \vec{f}_{ij}, \label{newton}
\end{align}
\end{subequations}
in three dimensiones, where $\svec{r}_i$ is the position, $\svec{\alpha}_i$ is the angle,
$\vec{p}_i$ the momentum and $\svec{\phi}_i$ the angular momentum of particle $i$.

To define the Hamiltonian system, we pair these generalized position and momentum vectors as follows
\[
\vec{Q}(t)=\{\vec{r}_i(t),\svec{\alpha}_i(t)\}_{i=1}^N,\quad
\vec{P}(t)=\{\vec{p}_i(t),\svec{\phi}_i(t)\}_{i=1}^N.
\]

Then the kinetic energy can be calculated using only the generalized momenta $\vec{P}$ as follows
\[
T(\vec{P})=\sum_{i=1}^N \left( \frac{|\vec{p}_i|^2}{2m_i}+ \frac{|\svec{\phi}_i|^2}{2I_i}\right).\
\]
The potential energy is a combination of the potential of gravity,
the potential of the normal and tangential springs and the internal potential energy in the particles,
created from the remaining potential of the tangential spring at the time $t_{ij}^e$
that a particle pair $\{i,j\}$ looses contact,
\[ \{t_{ij}^e\}=\{t:\ \frac{d_i+d_j}{2}-|\vec{r}_{ij}(t)|=0,\ \frac{d}{dt}|\vec{r}_{ij}(t)|>0\}\]
with $\vec{r}_{ij}=\vec{r}_j(t)-\vec{r}_i(t)$.
The potential can be expressed in terms of the position and the tangential springs at all times, which itself is a
function of the previous positions of the particle pair,
\begin{subequations}\label{eq:Ham}
\begin{eqnarray}
V(\vec{Q})&=& V_{grav}(\vec{Q})+V_{ela}(\vec{Q})+V_{int}(\vec{Q}).
\end{eqnarray}
where the gravitational, elastic and internally stored potential energy is defined by
\begin{eqnarray}
V_{grav}(\vec{Q})&=& \sum_{i=1}^N -m_i \vec{r}_i\cdot\vec{g} 
\\
V_{ela}(\vec{Q})&=& \sum_{i=1}^N\sum_{j=i+1}^N \bigg( \frac{k^n}{2} |\pmb{\delta_{ij}}^n|^2 + \frac{k^{\,t}}{2} |\pmb{\delta}_{ij}^t|^2\bigg).
\\
V_{int}(\vec{Q})&=& \sum_{i=1}^N\sum_{j=i+1}^N \sum_{t_{ij}^e<t} \bigg( \frac{k^{\,t}}{2} |\pmb{\delta}_{ij}^t({t_{ij}^e}^-)|^2 \bigg).
\end{eqnarray}
\end{subequations}
We will now show that the total energy $H=T+V$ satisfies the Hamiltonian equations,
\begin{subequations}\label{eq:Ham}
\begin{gather}
\frac{\partial H}{\partial \vec{r}_i}  = -\frac{d\vec{p}_i}{dt},\quad
\frac{\partial H}{\partial \svec{\alpha}_i}  = -\frac{d\svec{\phi}_i}{dt}, \label{eq:Ham1}\\
\frac{\partial H}{\partial \vec{p}_i}  = \frac{d\vec{r}_i}{dt},\text{ and }
\frac{\partial H}{\partial \svec{\phi}_i}  = \frac{d\svec{\alpha}_i}{dt}.\label{eq:Ham2}
\end{gather}
\end{subequations}

%
To derive  \eqref{eq:Ham1}, we calculate 
\[\frac{\partial H}{\partial \vec{r}_i} = \frac{\partial}{\partial \vec{r}_i} \left(-m_i \vec{r}_i\cdot\vec{g} 
+\sum_{j\not=i} \left(\frac{k^n}{2} |\pmb{\delta}_{ij}^n|^2 + \frac{k^{\,t}}{2} |\pmb{\delta}_{ij}^t|^2 \right)\right) \label{dr_H}\]
term by term.
We can show that
\begin{eqnarray}\nonumber
\frac{\partial}{\partial \vec{r}_i} \frac{k^n}{2} {\delta_{ij}^n}^2
&=& k^n \max\left(0,(d_i+d_j)/2-r_{ij} \right) \frac{\partial \max\left(0,(d_i+d_j)/2-r_{ij} \right)}{\partial \vec{r}_i}
\\\nonumber&=& k^n \max\left(0,(d_i+d_j)/2-r_{ij} \right) \frac{\partial \left((d_i+d_j)/2-r_{ij} \right)}{\partial \vec{r}_i} 
\\\nonumber&=& -k^n \max\left(0,(d_i+d_j)/2-r_{ij} \right) \frac{\vec{r}_{ij}}{r_{ij}}
\\&=& -k^n \delta_{ij}^n \vec{n}_{ij}, \label{dr_deltan2}
\end{eqnarray}
where we used the fact that $\max\left(0,(d_i+d_j)/2-r_{ij} \right)$ is continuous in time.

Further, we take the derivative $\partial_{r_{i a}} \pmb{\delta}_{ij}^t$ ($a$ denotes the vector coordinate)
by the chain rule and use \eqref{eq:deltat} to show that
\begin{eqnarray} \nonumber
\frac{\partial \pmb{\delta}_{ij}^t}{\partial r_{i a}(t)}
&=& \frac{\partial t}{\partial r_{i a}(t)}\ \cdot 
\left.\frac{\partial \pmb{\delta}_{ij}^t}{\partial t}\right|_{\vec{v}_j=\svec{\omega}_i=\svec{\omega}_j=\vec{0},\ v_{i b}=0,b\not=a}
\\\nonumber&=& \frac{1}{v_{i a}} \cdot \left( v_{i a} \svec{\epsilon}_a + (-v_{i a} \svec{\epsilon}_a \cdot\vec{n}_{ij})\vec{n}_{ij} 
+ (\pmb{\delta}_{ij}^t\cdot(-v_{i a} \svec{\epsilon}_a))\frac{\vec{r}_{ij}}{\vec{r}_{ij}^2} \right) 
\\&=& \svec{\epsilon}_a - n_{ij a} \vec{n}_{ij} 
- \delta_{ij a}^t\frac{\vec{r}_{ij}}{\vec{r}_{ij}^2}, \label{dr_deltat}
\end{eqnarray} 
where $\svec{\epsilon}_a$ denotes the $a$-th basis vector of the coordinate system.
Here, \eqref{dr_deltat} becomes
\begin{eqnarray} \nonumber
\frac{\partial}{\partial r_{i a}} \frac{k^{\,t}}{2} {\pmb{\delta}_{ij}^t}^2
&=& k^{\,t} \pmb{\delta}_{ij}^t \cdot \frac{\partial \pmb{\delta}_{ij}^t}{\partial r_{i a}}
\\\nonumber&=& k^{\,t} \pmb{\delta}_{ij}^t \cdot \left(\svec{\epsilon}_a - n_{ij a} \vec{n}_{ij} 
- \delta_{ij a}^t\frac{\vec{r}_{ij}}{\vec{r}_{ij}^2}\right)
\\&=& k^{\,t} \pmb{\delta}_{ij}^t \cdot \svec{\epsilon}_a = k^{\,t} \delta_{ij a}^t, \label{dr_deltat2}
\end{eqnarray}
where the cancellation of terms arises because the tangential spring is orthogonal to the normal vector.
After substituting \eqref{dr_deltat2} and \eqref{dr_deltan2} into \eqref{dr_H} we obtain
\begin{eqnarray} \nonumber
\frac{\partial H}{\partial \vec{r}_i} 
&=& \frac{\partial}{\partial \vec{r}_i} (-m_i \vec{r}_i\cdot\vec{g} 
+\sum_{j\not=i} \frac{k^n}{2} |\pmb{\delta_{ij}^n}|^2 + \frac{k^{\,t}}{2} {|\pmb{\delta}_{ij}^t}|^2).
\\&=& -m_i \cdot\vec{g} 
-\sum_{j\not=i} k^n \delta_{ij}^n \vec{n}_{ij} - k^{\,t} \pmb{\delta}_{ij}^t
~=~ -\frac{d \vec{p}_i}{dt}.
\end{eqnarray} 

\medskip
%
Next, we calculate
\[
\frac{\partial H}{\partial \alpha_{i a}} = 
\frac{\partial}{\partial \alpha_{i a}} \sum_{j\not=i} k^{\,t}/2 |\pmb{\delta}_{ij}^t|^2 
~=~ \sum_{j\not=i} k^{\,t} \pmb{\delta}_{ij}^t \cdot \frac{\partial}{\partial \svec{\alpha}_{i a}} \pmb{\delta}_{ij}^t. \label{da_H}
\]
We take the derivative $\partial_{\alpha_{i a}(t)} \pmb{\delta}_{ij}^t$  using the chain rule
and equations \eqref{eq:deltat} and \eqref{eq:vt}
\begin{eqnarray} \nonumber
\frac{\partial \pmb{\delta}_{ij}^t}{\partial \alpha_{i a}(t)}
&=& \frac{\partial t}{\partial \alpha_{i a}(t)}\ 
\left.\frac{\partial \pmb{\delta}_{ij}^t}{\partial t}\right|_{\vec{v}_i=\vec{v}_j=\svec{\omega}_j=\vec{0},\ \omega_{i b}=0,b\not=a}
\\\nonumber&=&  \frac{\partial t}{\partial \alpha_{i a}(t)}\ 
\left.\frac{\partial (-\omega_i \svec{\epsilon}_a \times \vec{b}_{ij})}{\partial t}\right|_{\vec{v}_i=
\vec{v}_j=\svec{\omega}_j=\vec{0},\ \omega_{i b}=0,b\not=a}
\\&=& -\svec{\epsilon}_a \times \vec{b}_{ij}.
\label{da_deltat}
\end{eqnarray} 
Substituting \eqref{da_deltat} into \eqref{da_H} we obtain
\begin{eqnarray} 
\frac{\partial H}{\partial \alpha_{i a}}
&=& -\sum_{j\not=i} k^{\,t} \pmb{\delta}_{ij}^t \cdot (\svec{\epsilon}_a \times \vec{b}_{ij})
~=~ \sum_{j\not=i} (k^{\,t}\pmb{\delta}_{ij}^t \times \vec{b}_{ij})_a,
\end{eqnarray} 
where we used the identity
\[\vec{c} \cdot(\svec{\epsilon}_a \times \vec{d}) = -(\vec{c}\times\vec{d})_a\ \forall \vec{c},\vec{d}\in\mathbb{R}^3.\]
Thus, using \eqref{da_H} and that $\vec{b}_{ij}$ and $\vec{n}_{ij}$ are parallel, we obtain
\begin{eqnarray} 
\frac{\partial H}{\partial \svec{\alpha}_i}
&=& -\sum_{j\not=i} (\vec{b}_{ij} \times k^{\,t}\pmb{\delta}_{ij}^t)
~=~ \sum_{j\not=i} (\vec{b}_{ij} \times \vec{f}_{ij})
~=~ -\frac{d \svec{\phi}_i}{dt}.
\end{eqnarray} 

Subsequently, we derive \eqref{eq:Ham2} since
\begin{equation} 
\frac{\partial H}{\partial \vec{p}_i} 
= \frac{\partial}{\partial \vec{p}_i} \frac{|\vec{p_i}|^2}{2m_i}= \frac{\vec{p_i}}{m_i} = \frac{d \vec{r}_i}{dt}, 
\end{equation}
and
\begin{equation} 
\frac{\partial H}{\partial \svec{\phi}_i} 
= \frac{\partial}{\partial \svec{\phi}_i} \frac{\vec{\phi_i}^2}{2I_i}= \frac{\vec{\phi_i}}{I_i} = \frac{d \svec{\alpha}_i}{dt}. 
\end{equation}

Finally, we show that the total energy is conserved. Since mass $m_i$, radius $r_i$ and spring constants $k^n$,
$k^{\,t}$ are constant, $H$ has no direct dependence on $t$ and thus 
\begin{equation}\label{eq:dtH}
\frac{\partial H}{\partial t}=0.
\end{equation}
Using and \eqref{eq:Ham} and \eqref{eq:dtH} yields
\begin{eqnarray} \nonumber
\frac{d}{dt} H(t,\vec{r},\vec{P})
&=& \frac{\partial H}{\partial t}
+ \sum_{i=1}^N \frac{\partial H}{\partial \vec{r}_i}\cdot\frac{d \vec{r}_i}{dt} +
\frac{\partial H}{\partial\svec{\alpha}_i}\cdot\frac{d \svec{\alpha}_i}{dt}
+ \frac{\partial H}{\partial \vec{p}_i}\cdot\frac{d \vec{p}_i}{dt} +
\frac{\partial H}{\partial \svec{\phi}_i}\cdot\frac{d \svec{\phi}_i}{dt}
\\\nonumber&=& \frac{\partial H}{\partial t}
- \frac{d \vec{p}_i}{dt} \cdot\frac{d \vec{r}_i}{dt}
- \frac{d \svec{\phi}_i}{dt} \cdot\frac{d \svec{\alpha}_i}{dt}
+ \frac{d \vec{r}_i}{dt} \cdot\frac{d \vec{p}_i}{dt}
+ \frac{d \svec{\alpha}_i}{dt} \cdot\frac{d \svec{\phi}_i}{dt}
\\&=& 0,
\end{eqnarray}

\begin{figure}[htbp]
%
\begin{psfrags}%
\psfragscanon%
%
\psfrag{s01}[t][t]{\fontsize{10}{15}\fontseries{m}\mathversion{normal}\fontshape{n}\selectfont \color[rgb]{0,0,0}\setlength{\tabcolsep}{0pt}\begin{tabular}{c}$t$\end{tabular}}%
\psfrag{s05}[l][l]{\fontsize{10}{15}\fontseries{m}\mathversion{normal}\fontshape{n}\selectfont \color[rgb]{0,0,0}$T+V$}%
\psfrag{s06}[l][l]{\fontsize{10}{15}\fontseries{m}\mathversion{normal}\fontshape{n}\selectfont \color[rgb]{0,0,0}$T$}%
\psfrag{s07}[l][l]{\fontsize{10}{15}\fontseries{m}\mathversion{normal}\fontshape{n}\selectfont \color[rgb]{0,0,0}$V_{ela}$}%
\psfrag{s08}[l][l]{\fontsize{10}{15}\fontseries{m}\mathversion{normal}\fontshape{n}\selectfont \color[rgb]{0,0,0}$V_{int}$}%
\psfrag{s09}[l][l]{\fontsize{10}{15}\fontseries{m}\mathversion{normal}\fontshape{n}\selectfont \color[rgb]{0,0,0}$T+V$}%
\psfrag{s11}[][]{\fontsize{10}{15}\fontseries{m}\mathversion{normal}\fontshape{n}\selectfont \color[rgb]{0,0,0}\setlength{\tabcolsep}{0pt}\begin{tabular}{c} \end{tabular}}%
\psfrag{s12}[][]{\fontsize{10}{15}\fontseries{m}\mathversion{normal}\fontshape{n}\selectfont \color[rgb]{0,0,0}\setlength{\tabcolsep}{0pt}\begin{tabular}{c} \end{tabular}}%
%
\fontsize{10}{15}\fontseries{m}\mathversion{normal}%
\fontshape{n}\selectfont%
%
\psfrag{x01}[t][t]{0}%
\psfrag{x02}[t][t]{0.2}%
\psfrag{x03}[t][t]{0.4}%
\psfrag{x04}[t][t]{0.6}%
\psfrag{x05}[t][t]{0.8}%
\psfrag{x06}[t][t]{1}%
\psfrag{x07}[t][t]{1.2}%
\psfrag{x08}[t][t]{1.4}%
\psfrag{x09}[t][t]{1.6}%
%
\psfrag{v01}[r][r]{0}%
\psfrag{v02}[r][r]{1}%
\psfrag{v03}[r][r]{2}%
\psfrag{v04}[r][r]{3}%
\psfrag{v05}[r][r]{4}%
\psfrag{v06}[r][r]{5}%
\psfrag{v07}[r][r]{6}%
%
\resizebox{8cm}{!}{\includegraphics{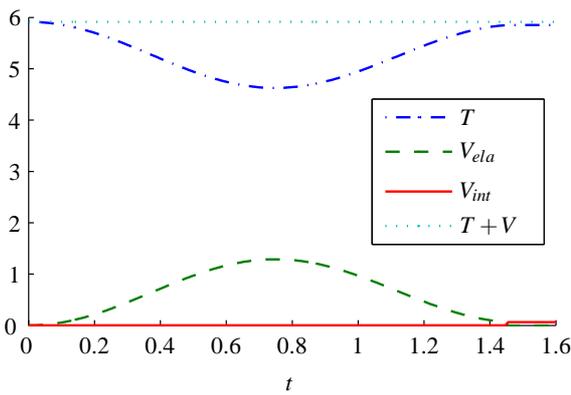}}%
\end{psfrags}%
%

\caption{Energy balance of two particles colliding non-collinear in the dissipation- and yield-free case.
The kinetic and elastic potential energy are in balance, until the particles loose contact and potential
spring energy is converted into internal energy.}
\label{fig:HamiltonianSystem}
\end{figure}

Fig.~\ref{fig:HamiltonianSystem} shows the energy balance of two particles colliding non-collinear
in the dissipation- and yield-free case. One can see the jump in energy at the end of contact,
where potential tangential spring energy is converted into internal energy.



\section{Orthogonality of the tangential spring} \label{sec:tangentialspring}

To show that the tangential spring is orthogonal to $\vec{r}_{ij}$,
note that the tangential spring is initially of zero length and therefore orthogonal to $\vec{r}_{ij}$; further
\begin{eqnarray}\nonumber
\frac{d (\svec{\delta}_{ij}^t\cdot \vec{r}_{ij})}{dt} &=&
\frac{d \svec{\delta}_{ij}^t}{dt}\cdot \vec{r}_{ij} +
\svec{\delta}_{ij}^t \cdot \vec{v}_{ij}
\\\nonumber
&=& \left(\vec{v}_{ij}^t-\frac{\svec{(\delta}_{ij}^t\cdot \vec{v}_{ij})\vec{r}_{ij}}{r_{ij}^2}
\right) \cdot \vec{r}_{ij} + \svec{\delta}_{ij}^t \cdot \vec{v}_{ij}
\\
&=& \vec{v}_{ij}^t \cdot \vec{r}_{ij}
= 0 \label{eq:orth}
\end{eqnarray}
Thus, we can integrate equation \eqref{eq:orth} to obtain a continuously orthogonal tangential spring with
$\svec{\delta}_{ij}^t\cdot \vec{r}_{ij}=0$. 



\section{Algorithms for time integration and the calculation of the tangential force}\label{sec:ti}\label{sec:ft}
The algorithm for the time integration and the calculation of the tangential force is shown in Algorithms \ref{alg:timeint} and \ref{alg:tangforce}.

\begin{algorithm}
\DontPrintSemicolon 
\KwData{Initial positions and translational and angular velocities $\vec{r}_i^0,\vec{v}_i^0,\vec{\omega}_i^0$, masses $m_i$, inertias $I_i$, time step $\Delta t$} 
$\vec{r}_i \longleftarrow \vec{r}_i^0$,\ $\vec{v}_i\longleftarrow \vec{v}_i^0$,\ $\vec{\omega}_i\longleftarrow \vec{\omega}_i^0\ \forall i$\;
$(\vec{f}_i,\vec{q}_i) \longleftarrow \text{forces-and-torques}(\{\vec{r}_j, \vec{v}_j,\vec{\omega}_j\}_{j=1}^N)\ \forall i$\;
\For{$i=1,2,\dots,N$}{
	$\vec{v}_i \longleftarrow \vec{v}_i+\frac{\Delta t}{2} \frac{\vec{f}_i}{m_i}\ \forall i$\;
	$\vec{\omega}_i \longleftarrow \vec{\omega}_i+\frac{\Delta t}{2} \frac{\vec{q}_i}{I_i}\ \forall i$\;
	$\vec{r}_i \longleftarrow \vec{r}_i+\Delta t \vec{v}_i\ \forall i$\;
	\ForEach{particle pair $(i, j)$ in contact}{
		\lIf{contact is new}{$\vec{\delta}_{ij}^t \longleftarrow \vec{0}$}\;
		$\vec{a}_{ij}^t \longleftarrow \vec{v}_{ij}^t-\frac{\svec{(\delta}_{ij}^t\cdot \vec{v}_{ij})\vec{r}_{ij}}{r_{ij}^2}$\;
		$\vec{\delta}_{ij}^t \longleftarrow \vec{\delta}_{ij}^t + \Delta t \vec{a}_{ij}^t$
	}
	$(\vec{f}_i,\vec{q}_i) \longleftarrow \text{forces-and-torques}(\{\vec{r}_j, \vec{v}_j,\vec{\omega}_j\}_{j=1}^N)\ \forall i$\;
	$\vec{v}_i \longleftarrow \vec{v}_i+\frac{\Delta t}{2} \frac{\vec{f}_i}{m_i}\ \forall i$\;
	$\vec{\omega}_i \longleftarrow \vec{\omega}_i+\frac{\Delta t}{2} \frac{\vec{q}_i}{I_i}\ \forall i$\;
}
\caption{Time integration}
\label{alg:timeint}
\end{algorithm}



\begin{algorithm}
\DontPrintSemicolon 
$\vec{f}_{ij}^{t} \longleftarrow -k^t \svec{\delta}_{ij}^t -\gamma^t \vec{v}_{ij}^t$\;
\If{$(|\vec{f}_{ij}^{t}|>\mu_c|\vec{f}_{ij}^{n}|)$}{
	$\vec{f}_{ij}^{t} \longleftarrow \mu_c \frac{|\vec{f}_{ij}^{n}|}{|\vec{f}_{ij}^{t}|} \vec{f}_{ij}^{t}$\;
	$\delta_{ij}^t \longleftarrow -\frac{1}{k^t}(\vec{f}_{ij}^{t}+\gamma^t \vec{v}_{ij}^t)$
}
\caption{Calculation of the tangential force, including sliding}
\label{alg:tangforce}
\end{algorithm}

\end{document}